\documentclass[11pt, letterpaper,  onecolumn, draftcls]{IEEEtran}
\usepackage{graphicx}
\usepackage{amsmath}
\usepackage{amssymb}
\usepackage{graphics}
\usepackage{latexsym}
\usepackage[dvips]{epsfig}
\usepackage{cite}
\usepackage{subfigure}
\usepackage{setspace}
\usepackage{tabularx}

\newtheorem{Definition}{Definition}
\newtheorem{Lemma}{Lemma}

\newtheorem{Proposition}[Lemma]{Proposition}
\newtheorem{Theorem}{Theorem}

\newtheorem{Example}{Example}
\newtheorem{Remark}{Remark}

\def\Pr{{\rm {Pr}}}
\def\E{{\rm  E}}
\def\Var{{\rm {Var}}}

\allowdisplaybreaks[1]

\ifCLASSINFOpdf
\else
\fi
\begin{document}
%
\title{\huge Cut-Set Bounds for Multimessage Multicast Networks with Independent Channels and Zero-Delay Edges}


\author{Silas~L.~Fong, \IEEEmembership{Member,~IEEE}
\thanks{S.~L.~Fong is with the Department of Electrical and Computer Engineering, National University of Singapore (NUS), Singapore 117583 (e-mail: \texttt{silas\_fong@nus.edu.sg}).}
\thanks{This paper was presented in part at 2015 IEEE International Symposium on Information Theory, Hong Kong.}
}%


%


\maketitle

\begin{abstract}
We consider a communication network consisting of nodes and directed edges that connect the nodes. The network may contain cycles.
The communications are slotted where the duration of each time slot is equal to the maximum propagation delay experienced by the edges.
The edges with negligible delays are allowed to be operated before the other edges in each time slot. For any pair of adjacent edges $(\ell, i)$ and $(i,j)$ where~$(\ell,i)$ terminates at node~$i$ and~$(i,j)$ originates from node~$i$, we say $(\ell,i)$ \textit{incurs zero delay} on $(i,j)$ if~$(\ell,i)$ is operated before~$(i,j)$; otherwise, we say $(\ell,i)$ \textit{incurs a unit delay} on $(i,j)$. In the classical model, every edge incurs a unit delay on every adjacent edge and the cut-set bound is a well-known outer bound on the capacity region.
In this paper, we investigate the multimessage multicast network (MMN) consisting of independent channels where each channel is associated with a set of edges and each edge may incur zero delay on some other edges.
Our result reveals that the capacity region of the MMN with independent channels and zero-delay edges lies within the classical cut-set bound despite a violation of the unit-delay assumption.
%
\end{abstract}



%
\IEEEpeerreviewmaketitle

\section{Introduction}\label{introduction}
\IEEEPARstart{T}{his} paper studies time-slotted communications over networks consisting of nodes and directed edges that connect the nodes, and the networks may contain cycles. Each edge receives a symbol from a node and outputs a symbol to a node in each time slot, where the duration of a time slot is set to be the maximum propagation delays experienced by the edges. In practical communication networks, propagation delays of different links may vary significantly due to different distances and different transmission medium (e.g., optical fiber, air, water, etc.) across different links. For example, links with relatively short distances have shorter propagation delays compared to those with relatively long distances, and links established through the optical fiber medium generally experience negligible propagation delays compared to links established through the water medium. In order to characterize the scenario where the propagation delays experienced by some edges are negligible compared to the delays experienced by the other edges, we allow the edges with negligible delays to be operated before the rest of the edges in each time slot. Since the symbols transmitted on earlier-operated edges may depend on the symbols output from latter-operated edges, we say that an edge $(\ell,i)$ terminating at node~$i$ \textit{incurs zero delay} on an edge $(i,j)$ originating from node~$i$ if~$(\ell,i)$ is operated before~$(i,j)$; otherwise, we say $(\ell,i)$ \textit{incurs a unit delay} on $(i,j)$.
%
Similarly, the network is said to \textit{contain zero-delay edges} if there exists an edge that incurs zero delay on another edge; the network is said to \textit{contain no zero-delay edge} if every edge incurs a delay on every other edge. Under the classical model, every discrete memoryless network (DMN) \cite[Ch.~15]{CoverBook} is assumed to contain no zero-delay edge because all the edges are operated at the same time. A well-known outer bound on the capacity region of the DMN that contains no zero-delay edge is the {\em classical cut-set bound}~\cite{elgamal_81}. It is easy to construct a network with zero-delay edges whose capacity region is strictly larger than the cut-set bound. One such network is the binary symmetric channel with correlated feedback (BSC-CF) considered in \cite[Sec.~VII]{FongYeung15}, which will be introduced in the next subsection.

\subsection{Two Motivating Examples} \label{sectionMotivating}
\subsubsection{A two-way channel} 
\begin{figure}[!t]
 \centering
  \subfigure[BSC-DMCF / BSC-CF]{\includegraphics[width=1.5 in, height=0.9 in, bb = 203 258 468 425, angle=0, clip=true]{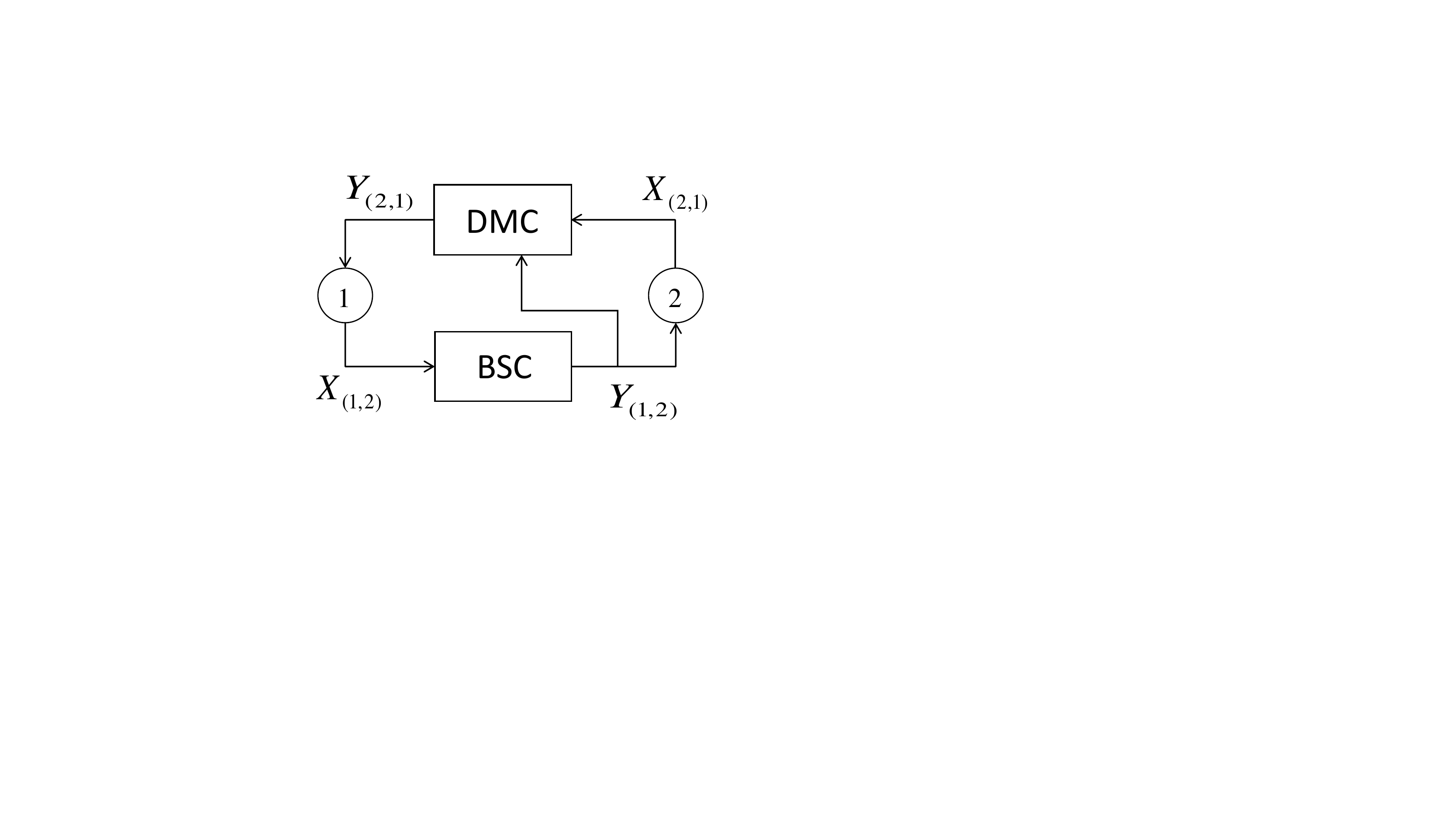}}
 \quad \qquad \subfigure[BSC-IF]{\includegraphics[width=1.5 in, height=0.9 in, bb = 203 258 468 425, angle=0, clip=true]{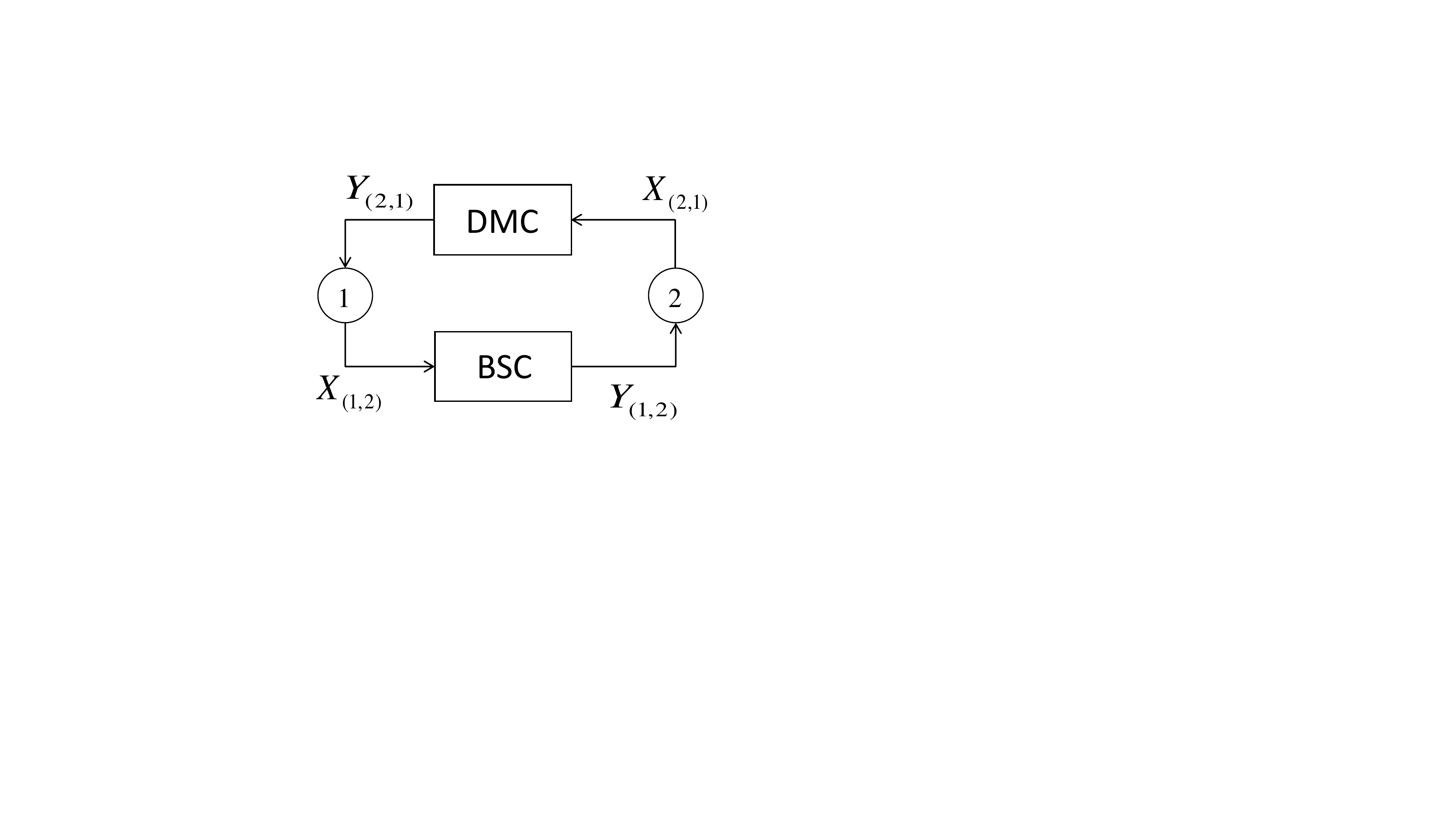}}
 \vspace{0.15 in}
\caption{Examples of the two-way channel.}  \label{BSCFB}
\end{figure}

Consider a network that consists of two nodes denoted by $1$ and $2$ respectively and two edges denoted by $(1,2)$ and $(2,1)$ respectively. Node~1 and node~2 want to transmit a message to each other. This is a two-way channel \cite{database:bib17}. In each time slot, node~1 and node~2 transmit $X_{(1,2)}$ and $X_{(2,1)}$ respectively, and they receive $Y_{(2,1)}$ and $Y_{(1,2)}$ respectively. All the input and output alphabets are binary, and the channel associated with edge~$(1,2)$ is a binary symmetric channel (BSC) while the channel associated with edge~$(2,1)$ is a discrete memoryless channel (DMC) whose output may depend on the output of channel~$(1,2)$. In this network, channel~$(1,2)$ incurs zero delay on channel~$(2,1)$, i.e., node~$2$ can receive $Y_{(1,2)}$ before encoding and transmitting $X_{(2,1)}$. We call this network the \textit{BSC with DMC feedback} (BSC-DMCF), which is illustrated in Figure~\ref{BSCFB}(a).

When $Y_{(2,1)}=X_{(2,1)}+Y_{(1,2)}$, the BSC-DMCF is also referred to as the BSC-CF in \cite[Sec.~VII]{FongYeung15}. It has been shown in \cite[Sec.~VII]{FongYeung15} that the capacity region of the BSC-CF is strictly larger than the classical cut-set bound, where the classical cut-set bound is obtained under the assumption that the network contains no zero-delay edge while the capacity region of the BSC-CF is achieved when edge~$(1,2)$ incurs zero delay on edge~$(2,1)$. Consequently, we have the following conclusion:
\begin{description}
\item[($\ast$)] \textit{The capacity region of some BSC-DMCF with zero-delay edges is strictly larger than the classical cut-set bound}.
\end{description}
However, Statement~($\ast$) is based on an important assumption: The outputs of the two channels can have correlation given their inputs. In other words, the noises of the two channels can be correlated given the channel inputs. If the noises are assumed to be independent given the channel inputs, i.e.,
\begin{align}
&\Pr\{Y_{(1,2)}=a, Y_{(2,1)}=b\,| X_{(1,2)}=c,X_{(2,1)}=d\} \notag\\
&=\Pr\{Y_{(1,2)}=a\,| X_{(1,2)}=c\}\Pr\{Y_{(2,1)}=b\, |X_{(2,1)}=d\} \label{eqnBSCIF}
 \end{align}
for all $(a,b,c,d)\in \{0,1\}^4$, then it is not clear whether Statement~($\ast$) still holds. To facilitate discussion, we call the BSC-DMCF which satisfies \eqref{eqnBSCIF} the \textit{BSC with independent feedback} (BSC-IF), which is illustrated in Figure~\ref{BSCFB}(b). Indeed, the BSC-IF with zero-delay edges always lies within the classical cut-set bound due to the fact that the two channels are independent and the well-known fact that the presence of instantaneous feedback does not increase the capacity of a point-to-point channel \cite[Sec.~7.12]{CoverBook}. Consequently, Statement~($\ast$) does not hold for the BSC-IF. This motivates us to investigate a general network with zero-delay edges under the assumption that the channels are independent, and compare its capacity region with the cut-set bound.

\subsubsection{A two-relay network (TRN)}\label{subsecTRN}
 \begin{figure}[!t]
 \centering
 \includegraphics[width=3 in, height=1.9 in, bb = 137 154 623 452, angle=0, clip=true]{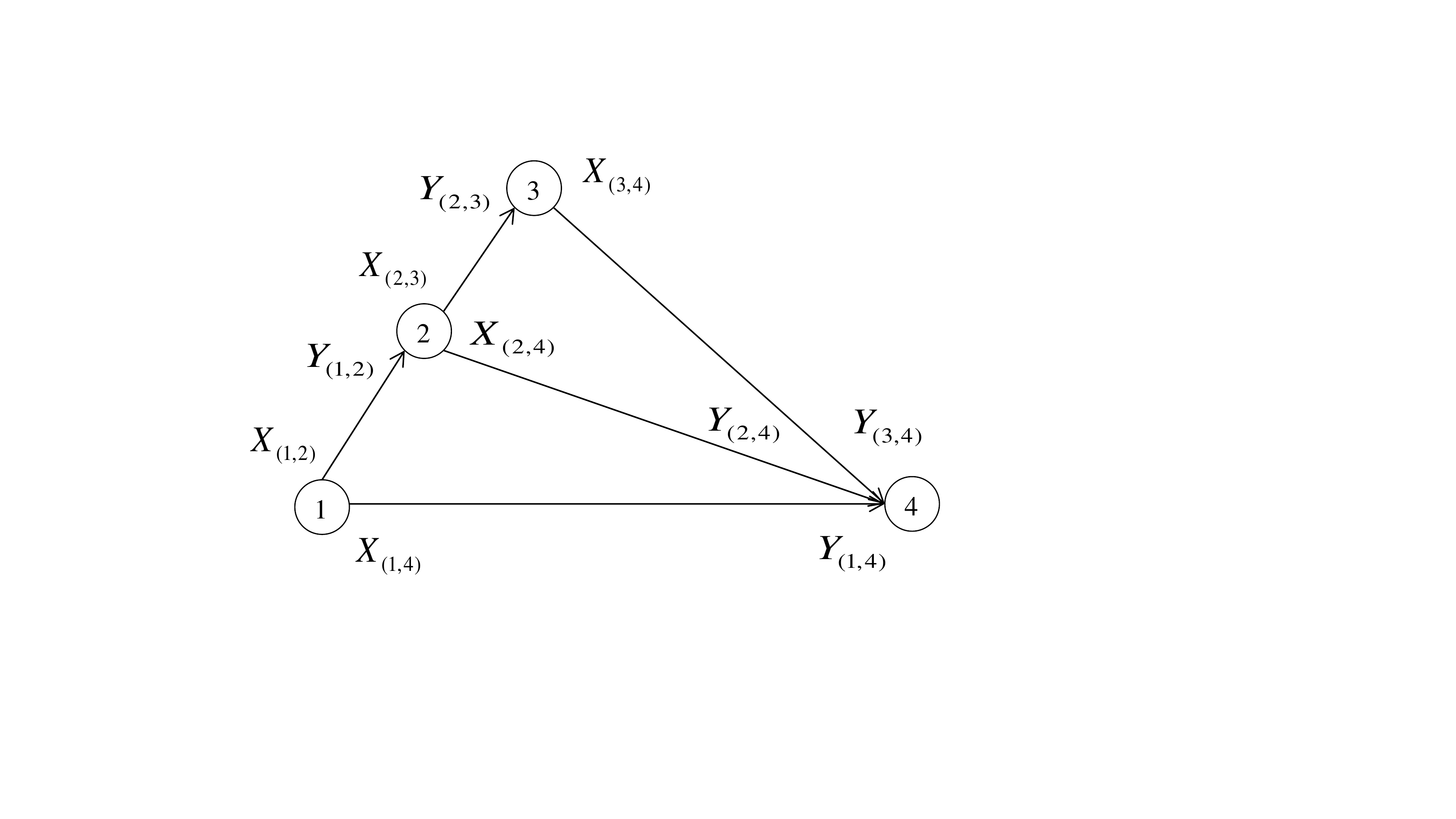}
\caption{TRN-CN / TRN-IN}  \label{2relayNetwork}
\end{figure}
Consider a two-relay network (TRN) illustrated in Figure~\ref{2relayNetwork}, which consists of a source denoted by~$1$, two relays denoted by~$2$ and~$3$ respectively, and a destination denoted by~$4$. The source wants to send information to the destination with the help of the two relays. Suppose all the input and output alphabets are binary. In each time slot, $X_{(i,j)}$ is transmitted on edge~$(i,j)$ by node~$i$ and $Y_{(i,j)}$ is received from edge~$(i,j)$ by node~$j$ for each edge~$(i,j)$ in the relay network. In addition, suppose
\begin{equation}
Y_{(1,2)}=U, \label{output2relayNetwork*}
\end{equation}
\begin{equation}
Y_{(2,3)}=V, \label{output2relayNetwork**}
\end{equation}
and
\begin{equation}
Y_{(1,4)}=Y_{(2,4)}=Y_{(3,4)}=X_{(1,4)}+X_{(2,4)}+X_{(3,4)}+U +V \label{output2relayNetwork}
\end{equation}
where $U$ and $V$ are two independent Bernoulli random variables with
\begin{equation}
\Pr\{U=0\}=\Pr\{V=0\}=1/2.
\end{equation}
 To facilitate discussion, we call the TRN described above the \emph{two-relay network with correlated noises (TRN-CN)}.
It can be easily seen that if edge~$(1,2)$ incurs zero delay on edge~$(2,4)$ and edge~$(2,3)$ incurs zero delay on edge~$(3,4)$, then node~4 can receive one bit per time slot from node~1 with the help of nodes~$2$ and~$3$ sending~$U$ and~$V$ respectively (cf.\ \eqref{output2relayNetwork}). On the contrary, if either edge~$(1,2)$ incurs a delay on edge~$(2,4)$ or edge~$(2,3)$ incurs a delay on edge~$(3,4)$, then node~4 cannot receive any information from node~1 because the independent uniform bits~$U$ and~$V$ cannot be completely cancelled simultaneously\footnote{As usual, the noises $U$ and $V$ generated in different time slots are assumed to be independent.}. Consequently, the capacity of the TRN-CN with zero-delay edges is strictly larger than the classical cut-set bound, i.e.,~$0$.

Consider another TRN with the same topology as illustrated in Figure~\ref{2relayNetwork} and specified by
\begin{equation}
Y_{(i,j)}=Z_{(i,j)}
\end{equation}
for all $(i,j)\ne (1,4)$
and
\begin{equation}
Y_{(1,4)}=X_{(1,4)}+X_{(2,4)}+X_{(3,4)}+Z_{(1,4)} \label{output2relayNetworkIN}
\end{equation}
where $Z_{(i,j)}$'s are independent Bernoulli random variables. To facilitate discussion, we call the TRN described above the \emph{two-relay network with independent noises (TRN-IN)}. It can be easily seen that the capacity of the TRN-IN, which is equal to the capacity of channel~$(1,4)$ specified in~\eqref{output2relayNetworkIN}, coincides with the cut-set bound even when zero-delay edges are present.

\subsection{Multimessage Multicast Network}
In this paper, we consider the multimessage multicast network (MMN) \cite[Ch.~18]{elgamal} consisting of independent channels, where the destination nodes want to decode the same set of messages transmitted by the source nodes. Two simple examples of the MMN with independent channel are the following two networks introduced in Section~\ref{sectionMotivating} --- the BSC-IF (where both nodes want to decode all the messages) and the TRN-IN, which belong to the class of MMNs consisting of independent discrete memoryless channels (DMCs) \cite{networkEquivalencePartI}. Note that the BSC-CF, unlike the BSC-IF, does not belong to the class of MMNs with independent DMCs because the forward and reverse channels of the BSC-CF are correlated (cf.\ Figure~\ref{BSCFB}(a)). Similarly, the TRN-CN, unlike the TRN-IN, does not belong to the class of MMNs with independent DMCs because the noises among the channels of the TRN-CN are correlated (cf.\ Figure~\ref{2relayNetwork}, \eqref{output2relayNetwork*}, \eqref{output2relayNetwork**} and \eqref{output2relayNetwork}).

 We propose an edge-delay model for the discrete memoryless MMN (DM-MMN) with independent channels and zero-delay edges. In our model, each channel is associated with a set of directed edges (e.g., a channel characterized by $q_{Y_{(1,4)}, Y_{(2,4)}|X_{(1,4)}, X_{(2,4)}}$ is associated with $\{(1,4), (2,4)\}$) and the channels are operated in a predetermined order so that the output random variables generated by earlier-operated channels are available for encoding the input random variables for latter-operated channels. Therefore, an edge may incur zero delay on another edge under our model. Our edge-delay model can be used to investigate the practical situation when some edges with negligible propagation delays are allowed to be operated before the rest of the edges in each time slot (for instance, in a cellular network, edges that connect the mobile users to their closest relays may experience negligible propagation delays compared to those that connect the relays to the base stations). The channels of the MMN are assumed to be \textit{independent}, meaning that the outputs among the channels are independent given their inputs, but the outputs within a channel are allowed to correlate with each other.
 \subsection{Main Contribution and Related Work} \label{subsecRelatedWork}
The main contribution of this paper is twofold: First, we establish an edge-delay model for the MMN consisting of independent channels which may contain zero-delay edges. Our model subsumes the classical model which assumes that every edge incurs a unit delay on every adjacent edge. Second, we prove that for each DM-MMN consisting of independent channels with zero-delay edges, the capacity region always lies within the classical cut-set bound despite a violation of the classical unit-delay assumption. Combining our cut-set bound result with existing achievability results from network equivalence theory~\cite{networkEquivalencePartI} and noisy network coding (NNC)~\cite{YAG11,Lim11}, we establish the tightness of our cut-set bound under our edge-delay model for the MMN with independent DMCs, and hence fully characterize the capacity region. More specifically, we show that the capacity region is the same as the set of achievable rate tuples under the classical unit-delay assumption. The capacity region result is then generalized to the MMN consisting of independent additive white Gaussian noise (AWGN) channels with zero-delay edges.

The MMN with independent DMCs has been investigated in~\cite{FongIT16} under the \textit{node-delay} model proposed in~\cite{FongYeung15} for general discrete memoryless networks. The precise definition of zero-delay nodes in~\cite[Sec.~IV]{FongYeung15} can be expressed under our edge-delay model as follows: A node \emph{incurs no delay} if and only if every incoming edge of the node incurs no delay on every outgoing edge of the node. In~\cite{FongIT16}, it was shown that the capacity region of the MMN with independent DMCs is equal to that under the classical unit-node-delay assumption even when zero-delay nodes are present. Under the node-delay model, all the outgoing edges of each zero-delay node must be operated simultaneously after all the incoming edges of the node have been operated (cf.\ \cite[Definitions~1 and~2]{FongIT16}) while under our edge-delay model, the incoming and outgoing edges of a node can be operated in some predetermined order (cf.\ Definition~\ref{defOrderedPartition}, \ref{defDiscreteNetwork} and~\ref{defChannelOperationSequence}). For example, for the TRN-CN described in Section~\ref{subsecTRN}, edge~$(2,3)$ can be operated before edge~$(2,4)$ under our edge-delay model but not under the node-delay model. In this work, we strengthen the main result in~\cite{FongIT16} under our edge-delay model for the MMN with independent DMCs and show that the capacity region remains unchanged even when zero-delay edges are present.

It was shown by Effros \cite{EffrosIndependentDelay} that under the positive-delay assumption\footnote{Effros's framework does not consider zero-delay edges, which can be seen from the encoding rules stated in \cite[Def.~1]{EffrosIndependentDelay} that assumes $X_{i,k}$ is a function of $(W_i, Y_i^{k-1})$.} in the classical setting, the set of achievable rate tuples for the MMN with independent channels does not depend on the amount of positive delay incurred by each edge on each other edge. Our capacity result for the MMN with independent DMCs (as well as AWGNs) complements Effros's finding as follows: The set of achievable rate tuples for the MMN with independent DMCs (as well as AWGNs) does not depend on the amount of delay incurred by each edge on each other edge, even with the presence of zero-delay edges. From a practical point of view, the capacity region of the MMN with independent DMCs (as well as AWGNs) is not affected by the way of handling delays among the channels or how the channels are synchronized, even when zero-delay edges are present.
%
\subsection{Paper Outline}
The rest of the paper is organized as follows.
Section~\ref{notation} presents the notation. Section~\ref{sectionFormulation} formulates our edge-delay model for the DM-MMN with independent channels and zero-delay edges and state our cut-set bound result, whose proof is contained in Section~\ref{sectionInnerOuterBound}. In Section~\ref{sectionMMNwithIndepDMCs}, we use our cut-set bound to prove the capacity region of the MMN with independent DMCs and zero-delay edges. Section~\ref{sectionGaussianChannels} generalizes our cut-set bound result to the MMN consisting of independent AWGN channels with zero-delay edges and characterizes its capacity region. Section~\ref{sectionConclusion} concludes this paper.

\section{Notation}\label{notation}
The sets of natural, real and non-negative real numbers are denoted by $\mathbb{N}$, $\mathbb{R}$ and $\mathbb{R}_+$ respectively. We use $(a)_+$ to denote $\max\{a, 0\}$.  We will take all logarithms to base 2. We use $\Pr\{E\}$ to represent the probability of an
event~$E$. We use an upper case letter~$X$ to denote a random variable with alphabet $\mathcal{X}$, and use a lower case letter $x$ to denote the realization of~$X$.
We use $X^n$ to denote a random tuple $(X_1,  X_2,  \ldots,  X_n)$, where the components $X_k$ have the same alphabet~$\mathcal{X}$.

For any arbitrary random variables~$X$ and~$Y$, we let $p_{X,Y}$ and $p_{Y|X}$ denote the probability distribution of $(X,Y)$ (can be both discrete, both continuous or one discrete and one continuous) and the conditional probability distribution of $Y$ given $X$ respectively.
We let $p_{X,Y}(x,y)$ and $p_{Y|X}(y|x)$ be the evaluations of $p_{X,Y}$ and $p_{Y|X}$ respectively at $(X,Y)=(x,y)$. To avoid confusion, we do not write $\Pr\{X=x, Y=y\}$ to represent $p_{X,Y}(x,y)$ unless $X$ and $Y$ are both discrete.
  We let $p_Xp_{Y|X}$ denote the joint distribution of $(X,Y)$, i.e., $p_Xp_{Y|X}(x,y)=p_X(x)p_{Y|X}(y|x)$ for all $x$ and $y$. If $X$ and $Y$ are independent, their joint distribution is simply $p_X p_Y$.  We let $\mathcal{N}(\,\cdot\, ;\mu,\sigma^2): \mathbb{R}\rightarrow [0,\infty)$ denote the probability density function of a Gaussian random variable whose mean and variance are $\mu$ and $\sigma^2$ respectively, i.e.,
$
\mathcal{N}(z;\mu,\sigma^2)\triangleq\frac{1}{\sqrt{2\pi \sigma^2}}e^{-\frac{(z-\mu)^2}{2\sigma^2}}$.

For any random tuple $(X,Y,Z)$ distributed according to $p_{X,Y,Z}$, we let $H_{p_{X,Z}}(X|Z)$ and $I_{p_{X,Y,Z}}(X;Y|Z)$ be the entropy of $X$ given $Z$ and mutual information between $X$ and $Y$ given $Z$ respectively. If $X$ is continuous, we let $h_{p_{X,Z}}(X|Z)$ be the differential entropy of~$X$ given~$Z$. If $X$, $Y$ and $Z$ are distributed according to $p_{X,Y,Z}$ and they form a Markov chain, we write
$(X\rightarrow Y\rightarrow Z)_p$. For simplicity, we drop the subscript of a notation if there is no ambiguity.
The $\mathcal{L}_1$-distance between
two distributions $p_X$ and $q_X$ on the same discrete alphabet $\mathcal{X}$, denoted by $\|p_X-q_X\|_{\mathcal{L}_1}$, is defined as $\|p_X-q_X\|_{\mathcal{L}_1}\triangleq\sum_{x\in\mathcal{X}}|p_X(x)-q_X(x)|$.


\section{Multicast Networks Consisting of Independent Channels}\label{sectionFormulation}
A multimessage multicast network (MMN) consists of $N$ nodes and $N^2$ directed edges. The MMN may contain cycles. Let
$
\mathcal{I}\triangleq \{1, 2, \ldots, N\}
$
be the index set of the nodes, and let
$
\mathcal{E}\triangleq \mathcal{I}\times\mathcal{I}
$
be the index set of the edges. Let $\mathcal{V}\subseteq \mathcal{I}$ and $\mathcal{D}\subseteq \mathcal{I}$ be the sets of sources and destinations respectively, where each source in~$\mathcal{V}$ transmits one message and each destination in~$\mathcal{D}$ wants to decode all the messages transmitted by the sources in~$\mathcal{V}$. We call $(\mathcal{V}, \mathcal{D})$ the \textit{multicast demand} for the network. The sources in $\mathcal{V}$ transmit information to the destinations in $\mathcal{D}$ in $n$ time slots (channel uses) as follows.
Node~$i$ transmits message
$
W_{i}\in \{1, 2, \ldots, M_i\}
$
for each $i\in \mathcal{V}$ and node $j$ wants to decode $\{W_{i}: i\in \mathcal{V}\}$ for each $j \in\mathcal{D}$, where~$M_i$ denotes the size of~$W_i$. We assume that each message $W_{i}$ is uniformly distributed over $\{1, 2, \ldots, M_i\}$ and all the messages are independent. For each $k\in \{1, 2, \ldots, n\}$ and each $(i,j)\in \mathcal{E}$, node~$i$ transmits $\{X_{(i,\ell),k} \}_{\ell\in\mathcal{I}}$ and receives $\{Y_{(\ell,j),k}\}_{\ell\in\mathcal{I}}$ in the $k^{\text{th}}$ time slot, where $X_{(i, \ell),k}$ is the symbol transmitted on edge~$(i, \ell)$ by node~$i$ and $Y_{(\ell, j),k}$ is the symbol received from edge~$(\ell, j)$ by node~$j$. The alphabet sets of $X_{(i,j),k}$ and $Y_{(i,j),k}$ are denoted by $\mathcal{X}_{(i,j)}$ and $\mathcal{Y}_{(i,j)}$ respectively for each $(i,j)\in \mathcal{E}$, where $\mathcal{X}_{(i,j)}$ and $\mathcal{Y}_{(i,j)}$ do not depend on the time index~$k$. After~$n$ time slots, node~$j$ declares~$\hat W_{i,j}$ to be the
transmitted~$W_{i}$ based on $(W_{j},\{Y_{(\ell ,j)}^n\}_{\ell\in\mathcal{I}})$ for each $(i, j)\in \mathcal{V}\times \mathcal{D}$.

To simplify notation, we use the following conventions for each $T_1, T_2\subseteq \mathcal{I}$: We let
$
 W_{T_1}\triangleq(W_{i} : i\in T_1)
$
 be the subtuple of $(W_{1}, W_{2}, \ldots, W_{N})$, and let
$
\hat W_{T_1\times T_2}\triangleq(\hat W_{i,j} : (i,j)\in T_1\times T_2)
$
be the subtuple of $(\hat W_{1,1}, \hat W_{1,2}, \ldots, \hat W_{N,N})$.
For any $N^2$-dimensional random tuple
$(X_{(1,1)}, X_{(1,2)}, \ldots, X_{(N,N)})$,
 we let
\[
 X_{T_1\times T_2} \triangleq (X_{(i,j)} : (i,j)\in T_1\times T_2)
 \]
  be its subtuple.
Similarly, for any $k\in \{1, 2, \ldots, n\}$ and any random tuple
$(X_{(1,1),k}, X_{(1,2),k}, \ldots, X_{(N,N),k})$,
 we let
\begin{equation}
X_{T_1\times T_2,k}\triangleq(X_{(i,j),k} : (i,j)\in T_1\times T_2)  \label{conventionXset}
\end{equation}
 be its subtuple.
\smallskip
\begin{Definition}\label{defOrderedPartition}
An $\alpha$-dimensional tuple $(\Omega_1, \Omega_2, \ldots, \Omega_\alpha)$ consisting of non-empty subsets of $\mathcal{E}$ is called an \textit{$\alpha$-partition of $\mathcal{E}$} if $\cup_{h=1}^\alpha \Omega_h = \mathcal{E}$ and $\Omega_i\cap \Omega_j  = \emptyset$ for all $i\ne j$.
\end{Definition}
\smallskip

\begin{Definition} \label{defDiscreteNetwork}
The discrete network with independent channels consists of $N^2$ finite input sets
$\mathcal{X}_{(1,1)}, \linebreak \mathcal{X}_{(1,2)}, \ldots, \mathcal{X}_{(N,N)}$, $N^2$ finite output sets $\mathcal{Y}_{(1,1)}, \mathcal{Y}_{(1,2)}, \ldots, \mathcal{Y}_{(N,N)}$ and $\alpha$ channels characterized by conditional distributions $q_{Y_{\Omega_1}|X_{\Omega_1}}^{(1)}$, $q_{Y_{\Omega_2}|X_{\Omega_2 }}^{(2)},\ldots,q_{Y_{\Omega_\alpha}|X_{\Omega_\alpha}}^{(\alpha)}$, where
$
\boldsymbol{\Omega} \triangleq (\Omega_1, \Omega_2, \ldots \Omega_\alpha)
$
  is an $\alpha$-partition of $\mathcal{E}$. We call $\boldsymbol{\Omega}$ the \textit{edge partition} of the network.
   The discrete network is denoted by $(\mathcal{X}_{\mathcal{E}}, \mathcal{Y}_\mathcal{E}, \alpha, \boldsymbol{\Omega}, \boldsymbol q)$ where
 $
  \boldsymbol q \triangleq (q^{(1)}, q^{(2)}, \ldots, q^{(\alpha)})$.
\end{Definition}
\smallskip

Under our model, the discrete network is characterized by $\alpha$ channels denoted by  $q_{Y_{\Omega_1}|X_{\Omega_1}}^{(1)}$, $q_{Y_{\Omega_2}|X_{\Omega_2 }}^{(2)},\ldots, \linebreak q_{Y_{\Omega_\alpha}|X_{\Omega_\alpha}}^{(\alpha)}$ as defined in Definition~\ref{defDiscreteNetwork}. If $\alpha=1$, our model simplifies to the classical model which is characterized by only one channel denoted by $q_{Y_{\mathcal{E}}|X_{\mathcal{E}}}^{(1)}$. The following example illustrates our model when $\alpha=3$.
\smallskip
\begin{Example}\label{example1}
For the BSC-IF described in Section~\ref{sectionMotivating}, we have $\mathcal{I}=\{1,2\}$ and $\mathcal{E} = \{(1,1),(1, 2), (2, 1), \linebreak (2,2)\}$, where $\mathcal{X}_{(1,2)}=\mathcal{Y}_{(1,2)}=\mathcal{X}_{(2,1)}=\mathcal{Y}_{(2,1)}=\{0,1\}$. Without loss of generality, we assume that $\mathcal{X}_{(1,1)}=\mathcal{Y}_{(1,1)}=\mathcal{X}_{(2,2)}=\mathcal{Y}_{(2,2)}=\{0\}$.
  The edge partition is denoted by $(\Omega_1, \Omega_2, \Omega_3)$ where $\Omega_1=\{(1,2)\}$, $\Omega_2=\{(2,1)\}$ and $\Omega_3 = \{(1,1),(2,2)\}$. Since $\mathcal{Y}_{(1,1)}=\mathcal{Y}_{(2,2)}=\{0\}$, $q_{Y_{\Omega_3}|X_{\Omega_3}}^{(3)}$ is a trivial channel which can carry no information. The other two channels of the BSC-IF are denoted by $q_{Y_{\Omega_1}|X_{\Omega_1}}^{(1)}= q_{Y_{(1,2)}|X_{(1,2)}}^{(1)}$ and $q_{Y_{\Omega_2}|X_{\Omega_2}}^{(2)} = q_{Y_{(2,1)}|X_{(2,1)}}^{(2)}$ respectively, which correspond to the BSC and the DMC in Figure~\ref{BSCFB}(b) respectively. \hfill $\blacksquare$
\end{Example}
\smallskip
\begin{Definition} \label{defDelayProfile}
An \textit{edge-delay profile}, also called \textit{delay profile} for simplicity, is an $N^3$-dimensional tuple $(b_{(\ell,i, j)}: (\ell, i, j)\in \mathcal{I}^3)$ where $b_{(\ell, i ,j)} \in \{0, 1\}$ represents the amount of delay incurred by edge~$(\ell, i)$ on edge $(i,j)$. The delay profile is said to be \textit{positive} if it is the all-one tuple $\mathbf{1}\triangleq (1, 1, \ldots, 1)$.
\end{Definition}
\smallskip
Under the classical model, only the positive delay profile is considered, i.e., every edge incurs a unit delay on all the edges. In our model, some elements of the delay profile can be zero, which indicates that some edge can incur zero delay on some other edges. However, if the delay profile contains too many zeros, deadlock loops may occur which result in each node waiting for the other nodes to transmit first before encoding and transmitting its outgoing symbols. The following two definitions enable us to formally define ``good" delay profiles that do not induce deadlock loops.
%
\smallskip
\begin{Definition} \label{defChannelOperationSequence}
A \textit{channel operation sequence} for the discrete network $(\mathcal{X}_\mathcal{E}, \mathcal{Y}_\mathcal{E}, \alpha, \boldsymbol{\Omega}, \boldsymbol q)$ is some permutation of the tuple $(1,2, \ldots,\alpha)$. The set of all channel operation sequences (permutations of $(1, 2, \ldots, \alpha)$) is denoted by $\Pi$.
\end{Definition}

When we formally define a code on the discrete network later, a channel operation sequence $(\pi(1), \pi(2), \linebreak \ldots, \pi(\alpha))\in \Pi$ together with a delay profile $B=(b_{(\ell, i ,j)}: (\ell, i, j)\in \mathcal{I}^3)$ will be associated with the code, where the $\alpha$ channels are operated in the order
 \[
 q^{(\pi(1))},  q^{(\pi(2))}, \ldots,  q^{(\pi(\alpha))}
 \]
 and $b_{(\ell, i ,j)}$ represents the amount of delay incurred by edge~$(\ell, i)$ on edge~$(i,j)$. If $b_{(\ell, i ,j)}=0$, then node~$i$ receives $Y_{(\ell,i),k}$ \textit{before} encoding $X_{(i,j),k}$, and we say $(\ell,i)$ \textit{incurs zero delay} on $(i,j)$; if $b_{(\ell, i ,j)}=1$, then node~$i$ receives $Y_{(\ell,i),k}$ \textit{after} encoding $X_{(i,j),k}$, and we say $(\ell,i)$ \textit{incurs a delay} on $(i,j)$. Under the classical model, $\alpha=1$ and the channel operation sequence can only be the tuple $(1)$ (because there is only one channel), which implies that $B$ can only be positive, meaning that the amount of delay incurred between every pair of edges can only be positive. In contrast, under our edge-delay model, the $\alpha$ channels can be operated in different orders and some elements of $B$ can take $0$ as long as deadlock loops  do not occur (in a deadlock loop, every node wants to receive first and then encode, and hence no transmission can take place). Therefore our model is a generalization of the classical model. The essence of the following definition is to characterize delay profiles which will not induce deadlock loops in the network.
\smallskip
\begin{Definition} \label{defFeasible}
Let $(\mathcal{X}_\mathcal{E}, \mathcal{Y}_\mathcal{E}, \alpha, \boldsymbol{\Omega}, \boldsymbol q)$ be a discrete network, and let $\boldsymbol{\pi}\triangleq(\pi(1), \pi(2), \ldots, \pi(\alpha))$ be a channel operation sequence. For each $(i,j)\in \mathcal{E}$, let $t_{(i, j)}$ be the unique integer such that $(i,j) \in \Omega_{\pi(t_{(i, j)})}$. Then, an edge-delay profile $(b_{(\ell, i ,j)}: (\ell, i, j)\in \mathcal{I}^3)$ is said to be \textit{feasible with respect to $\boldsymbol{\pi}$} if the following holds for each $(\ell,i,j)\in \mathcal{I}^3$: If $b_{(\ell, i ,j)}=0$, then $t_{(\ell, i)} < t_{(i, j)}$.
\end{Definition}

Under the classical model, the only possible delay profile is the positive delay profile, which is feasible with respect to any channel operation sequence according to Definition~\ref{defFeasible}. The following example illustrates a delay profile for the BSC-IF which is feasible with respect to one channel operation sequence but not to another channel operation sequence.
\begin{Example}\label{example2}
For the BSC-IF as defined in Example~\ref{example1}, $\alpha=3$ because it is characterized by three channels. Since the third channel is trivial, we assume without loss of generality that it is always the last channel to be operated. Then, the two channel operation sequences of our interest are $(1,2,3)$ and $(2,1,3)$. Let
\[
b_{(\ell,i,j)} \triangleq \begin{cases} 0 & \text{if $(\ell,i,j)=(1,2,1)$,}\\1 & \text{otherwise} \end{cases}
\]
for all $(\ell,i,j)\in \mathcal{I}^3$, and let
$B_{(1,2,1)}\triangleq (b_{(\ell,i,j)}: (\ell,i,j)\in \mathcal{I}^3)$ be the delay profile where only $b_{(1,2,1)}$ is~$0$. Then, it follows from Definition~\ref{defFeasible} that $B_{(1,2,1)}$ is \textit{feasible with respect to the channel operation sequence $(1,2,3)$} but $B_{(1,2,1)}$ is \textit{not feasible with respect to $(2,1,3)$}. An interpretation for the feasibility of $B_{(1,2,1)}$ is given as follows: If the channel is operated according to the sequence $(1,2,3)$, i.e., $q_{Y_{(1,2)}|X_{(1,2)}}^{(1)}$ is operated before $q_{Y_{(2,1)}|X_{(2,1)}}^{(2)}$, then $Y_{(1,2)}$ will be generated before the activation of $q_{Y_{(2,1)}|X_{(2,1)}}^{(2)}$. Therefore, node~$2$ can receive $Y_{(1,2)}$ before encoding $X_{(2,1)}$, and we say $B_{(1,2,1)}$ is feasible with respect to $(1,2,3)$. If the channel is operated according to the sequence $(2,1,3)$, then node~$2$ will receive $Y_{(1,2)}$ after the encoding of $X_{(2,1)}$, and we say $B_{(1,2,1)}$ is not feasible with respect to $(2,1,3)$.
\end{Example}

We are ready to define codes that use the network $n$ times.
\begin{Definition} \label{defCode}
Let $(\mathcal{X}_\mathcal{E}, \mathcal{Y}_\mathcal{E}, \alpha, \boldsymbol{\Omega}, \boldsymbol q)$ be a discrete network. Let $(\mathcal{V}, \mathcal{D})$ be the multicast demand for the network. In addition, let $\boldsymbol{\pi}\in \Pi$ be a channel operation sequence, and let
$
B\triangleq (b_{(\ell, i, j)} : (\ell, i,j)\in \mathcal{I}^3)
$
 be a feasible delay profile with respect to $\boldsymbol{\pi}$.
 A $(\boldsymbol{\pi}, B, n, M_{\mathcal{I}})$-code, where $M_\mathcal{I}\triangleq (M_1, M_2, \ldots, M_N)$, for~$n$ uses of the network consists of the following:
\begin{enumerate}
\item A message set
$
\mathcal{W}_{i}\triangleq \{1, 2, \ldots, M_i\}
$
 at node~$i$ for each $i\in \mathcal{I}$, where $M_i=1$ for each $i\in \mathcal{V}^c$. Message $W_i$ is uniformly distributed on $\mathcal{W}_i$.

\item An encoding function
\[
f_{(i,j),k} : \mathcal{W}_i \times \mathcal{Y}_{(1, i)}^{k-b_{(1,i,j)}} \times \ldots \times \mathcal{Y}_{(N, i)}^{k-b_{(N,i,j)}} \rightarrow \mathcal{X}_{(i,j)}
\]
 for each $(i,j)\in \mathcal{E}$ and each $k\in\{1, 2, \ldots, n\}$, where $f_{(i,j),k}$ is the encoding function for~$X_{(i,j),k}$ at node~$i$ in the
$k^{\text{th}}$ time slot such that
$
X_{(i,j),k}=f_{(i,j),k} (W_{i},
(Y_{(\ell,i)}^{k-b_{(\ell, i, j)}}: \ell\in \mathcal{I}))$.

\item A decoding function
$
g_{i,j} : \mathcal{W}_{j}  \times
\mathcal{Y}_{\mathcal{I}\times \{j\}}^{n} \rightarrow \mathcal{W}_{i}
$
 for each $(i, j) \in \mathcal{V}\times \mathcal{D}$, where $g_{i,j}$ is the decoding function for $W_{i}$ at node~$j$ such that
$
 \hat W_{i, j} \triangleq g_{i,j}(W_{j}, Y_{\mathcal{I}\times\{j\}}^{n})$.
\end{enumerate}
\end{Definition}

Given a $(\boldsymbol{\pi}, B, n, M_{\mathcal{I}})$-code, it follows from Definition~\ref{defCode} that for each $(\ell, i, j)\in \mathcal{I}^3$, edge~$(\ell,i)$ incurs a delay on edge~$(i,j)$ if $b_{(\ell, i, j)}=1$. If $b_{(\ell, i, j)}=0$, edge~$(\ell,i)$ incurs zero delay on edge~$(i,j)$, i.e., for each $k\in \{1, 2, \ldots, n\}$, node~$i$ receives $Y_{(\ell,i),k}$ before encoding $X_{(i,j),k}$. The feasibility condition of $B$ in Definition~\ref{defFeasible} ensures that the operations of any $(\boldsymbol{\pi},B, n, M_{\mathcal{I}})$-code are well-defined for the subsequently defined discrete memoryless network; the associated coding scheme is described after the network is defined.
\begin{Definition}\label{defDiscreteMemoryless}
A discrete network $(\mathcal{X}_\mathcal{E}, \mathcal{Y}_\mathcal{E}, \alpha, \boldsymbol{\Omega}, \boldsymbol q)$, when used multiple times,
 is called a \textit{discrete memoryless multimessage multicast network (DM-MMN)} if the following holds for any
 channel operation sequence $\boldsymbol{\pi}=(\pi(1), \pi(2), \ldots, \pi(\alpha))$ and any $(\boldsymbol{\pi}, B, n, M_{\mathcal{I}})$-code:

Let $U^{k-1}\triangleq (W_{\mathcal{I}}, X_{\mathcal{E}}^{k-1}, Y_{\mathcal{E}}^{k-1})$ be the collection of random variables that are generated before the $k^{\text{th}}$ time slot. To simplify notation, let
\begin{equation}
\Omega_{\boldsymbol{\pi}}^h \triangleq \bigcup_{m=1}^h \Omega_{\pi(m)}. \label{defOmegaSuperscript}
\end{equation}
 Then, for each $k\in\{1, 2, \ldots, n\}$ and each $h\in \{1, 2, \ldots, \alpha\}$,
\begin{align}
& \Pr\{U^{k-1} = u^{k-1}, X_{\Omega_{\boldsymbol{\pi}}^h ,k} =x_{\Omega_{\boldsymbol{\pi}}^h ,k}, Y_{\Omega_{\boldsymbol{\pi}}^h ,k}=y_{\Omega_{\boldsymbol{\pi}}^h ,k} \} \notag\\*
 & \quad =  \Pr\{U^{k-1} = u^{k-1}, X_{\Omega_{\boldsymbol{\pi}}^h ,k} =x_{\Omega_{\boldsymbol{\pi}}^h ,k},  Y_{\Omega_{\boldsymbol{\pi}}^{h-1} ,k}= y_{\Omega_{\boldsymbol{\pi}}^{h-1} ,k} \}
  q_{Y_{\Omega_{\pi(h)}}|X_{\Omega_{\pi(h)}}}^{(\pi(h))}(y_{\Omega_{\pi(h)},k}|x_{\Omega_{\pi(h)},k}) \label{memorylessStatement}
\end{align}
for all $u^{k-1}$, $x_{\Omega_{\boldsymbol{\pi}}^h ,k}$ and $y_{\Omega_{\boldsymbol{\pi}}^h ,k}$.
\end{Definition}

Following the notation in Definition~\ref{defDiscreteMemoryless}, consider any $(\boldsymbol{\pi}, B, n, M_{\mathcal{I}})$-code on the DM-MMN. In the $k^{\text{th}}$ time slot, $X_{\mathcal{E},k}$ and $Y_{\mathcal{E},k}$ are generated in the order
\begin{equation*}
X_{\Omega_{\pi(1)},k}, Y_{\Omega_{\pi(1)},k}, X_{\Omega_{\pi(2)},k}, Y_{\Omega_{\pi(2)},k}, \ldots, X_{\Omega_{\pi(\alpha)},k}, Y_{\Omega_{\pi(\alpha)},k} 
\end{equation*}
by transmitting on the channels in this order $q^{(\pi(1))},q^{(\pi(2))}, \ldots, q^{(\pi(\alpha))}$ using the $(\boldsymbol{\pi},B, n, M_{\mathcal{I}})$-code (as prescribed in Definition~\ref{defCode}). Specifically, $X_{\Omega_{\pi(h)}, k}$ and channel $q^{(\pi(h))}$ together define $Y_{\Omega_{\pi(h)},k}$ for each $h\in\{1, 2, \ldots, \alpha\}$. In addition, the feasibility condition of $B$ in Definition~\ref{defFeasible} ensures that the encoding of $X_{\Omega_{\pi(h)}, k}$ specified in Definition~\ref{defCode} is well-defined because $X_{\Omega_{\pi(h)}, k}$ depends on only the symbols generated before its encoding, which is formally shown in the following proposition.
\begin{Proposition}\label{propositionXFunctionOfY}
Let $\boldsymbol{\pi}\triangleq(\pi(1), \pi(2), \ldots, \pi(\alpha))$ be a channel operation sequence. Fix any $(\boldsymbol{\pi}, B, n, M_{\mathcal{I}})$-code
and fix an $h\in\{1, 2, \ldots, \alpha\}$. Then, for each $(i,j) \in \Omega_{\pi(h)}$, $X_{(i,j),k}$ is a function of $(W_i, Y_{\mathcal{I}\times\{i\}}^{k-1}, \linebreak  Y_{(\mathcal{I}\times\{i\}) \cap \Omega_{\boldsymbol{\pi}}^{h-1} ,k})$
 for each $k\in\{1, 2, \ldots, n\}$.
\end{Proposition}
\begin{IEEEproof}
Let $B\triangleq(b_{(\ell,i,j)}:(\ell,i,j)\in \mathcal{I}^3)$ be a delay profile that is feasible with respect to $\boldsymbol{\pi}$ and fix a $(\boldsymbol{\pi}, B, n, M_{\mathcal{I}})$-code. Fix a $k\in\{1, 2, \ldots, n\}$ and an edge $(i,j) \in \Omega_{\pi(h)}$. By Definition~\ref{defCode}, $X_{(i,j),k}$ is a function of $(W_i, (Y_{(\ell,i)}^{k-b_{(\ell,i,j)}}:\ell\in\mathcal{I}))$. Therefore, it suffices to show that $(Y_{(\ell,i)}^{k-b_{(\ell,i,j)}}:\ell\in\mathcal{I})$ is a function of $(Y_{\mathcal{I}\times\{i\}}^{k-1},  Y_{(\mathcal{I}\times\{i\}) \cap \Omega_{\boldsymbol{\pi}}^{h-1} ,k})$. Since each $b_{(\ell, i ,j)}$ is binary, it remains to prove the following statement:
 For each $\bar \ell$ such that $b_{(\bar \ell, i ,j)}=0$, $Y_{(\bar \ell,i),k}$ is a function of $Y_{(\mathcal{I}\times\{i\}) \cap \Omega_{\boldsymbol{\pi}}^{h-1} ,k})$. To this end, we fix an~$\bar \ell$ that satisfies $b_{(\bar \ell, i ,j)}=0$. Let $t_{(\bar \ell, i)}$ and $t_{(i, j)}$ be the two unique integers such that $(\bar \ell, i) \in \Omega_{\pi(t_{(\bar \ell, i)})}$ and $(i,j)\in\Omega_{\pi(t_{(i, j)})}$. It then follows from Definition~\ref{defFeasible} that $t_{(\bar \ell, i)} < t_{(i, j)}$, which implies that $(\bar \ell,i) \in \Omega_{\boldsymbol{\pi}}^{h-1} $, which then implies that $Y_{(\bar \ell,i),k}$ is a function of $Y_{(\mathcal{I}\times\{i\}) \cap \Omega_{\boldsymbol{\pi}}^{h-1} ,k})$.
\end{IEEEproof}
\smallskip

Under the classical model, $\alpha=1$ (there is only one channel) and $X_{\mathcal{E},k}$ and $Y_{\mathcal{E},k}$ are generated in the order
\begin{equation*}
X_{\mathcal{E},k}, Y_{\mathcal{E},k} 
\end{equation*}
in each time slot $k$.
The only channel operation sequence is $(1)$ and it follows from Definition~\ref{defFeasible} the only feasible delay profile with respect to $(1)$ is the positive delay profile, which implies that every edge incurs a unit delay on all the edges. Consequently, the classical model for the DM-MMN is a special case of our edge-delay model when $\alpha=1$. We are ready to define the capacity region of the DM-MMN using the following three definitions.
\smallskip
\begin{Definition} \label{defErrorProbability}
For a $(\boldsymbol{\pi}, B, n, M_{\mathcal{I}})$-code, let \begin{equation}
P_{\text{err}}^{n}\triangleq
\Pr\big\{\cup_{(i,j)\in \mathcal{V}\times \mathcal{D}} \{\hat{W}_{i,j} \ne   W_{i } \}\big\}\label{defErrorProbabilitySt}\end{equation}
be the probability of decoding error.
\end{Definition}

\begin{Definition} \label{defAchievableRate}
Let $\boldsymbol{\pi}$ be a channel operation sequence and $B$ be a feasible delay profile with respect to $\boldsymbol{\pi}$. A rate tuple $(R_{1}, R_{2}, \ldots, R_{N})$, denoted by $R_{\mathcal{I}}$, is \textit{$(\boldsymbol{\pi},B)$-achievable} for the DM-MMN if there exists a sequence of $(\boldsymbol{\pi},B, n, M_{\mathcal{I}})$-codes such that
$
 \liminf\limits_{n\rightarrow \infty} \frac{\log M_{i}}{n} \ge R_{i}
$
for each $i\in\mathcal{I}$ and
$
 \lim\limits_{n\rightarrow \infty} P_{\text{err}}^{n} = 0$.
A rate tuple is said to be \textit{$\boldsymbol{\pi}$-achievable} if it is $(\boldsymbol{\pi},B)$-achievable for some $B$. A rate tuple is said to be \textit{achievable} if it is $\boldsymbol{\pi}$-achievable for some $\boldsymbol{\pi}$.
\end{Definition}

Without loss of generality, we assume that $M_{i}=1$ and $R_{i}=0$ for all $i\in \mathcal{V}^c$ in the rest of this paper.
\begin{Definition}\label{defCapacityRegion}
The \textit{$(\boldsymbol{\pi},B)$-capacity region}, denoted by $\mathcal{C}_{B}^{\boldsymbol{\pi}}$, of the DM-MMN is the set consisting of every $(\boldsymbol{\pi},B)$-achievable rate tuple. The \textit{$\boldsymbol{\pi}$-capacity region} $\mathcal{C}^{\boldsymbol{\pi}}$ is defined as
\[
\mathcal{C}^{\boldsymbol{\pi}}\triangleq \bigcup_{\substack{B: B\text{ is feasible with}\\\text{\quad respect to $\boldsymbol{\pi}$}}} \mathcal{C}_{B}^{\boldsymbol{\pi}}.
\]
The \textit{capacity region} $\mathcal{C}$ is defined as
\[
\mathcal{C}\triangleq \bigcup_{\boldsymbol{\pi}\in \Pi} \mathcal{C}^{\boldsymbol{\pi}}.
\]
\end{Definition}
The following theorem is our main result.
\begin{Theorem} \label{thmCapacityRegionMulticast}
 Let $(\mathcal{X}_\mathcal{E}, \mathcal{Y}_\mathcal{E}, \alpha, \boldsymbol{\Omega}, \boldsymbol q)$ be a DM-MMN with independent channels and zero-delay edges under multicast demand $(\mathcal{V}, \mathcal{D})$, and let
 \begin{align}
\mathcal{R}_{\text{out}} \triangleq \!\!\!\!\! \bigcup_{\substack{p_{X_{\mathcal{E}}, Y_{\mathcal{E}}}:  p_{X_{\mathcal{E}}, Y_{\mathcal{E}}} = \\
\prod\limits_{h=1}^\alpha \left(p_{X_{\Omega_h}}
q^{(h)}_{Y_{\Omega_h} | X_{\Omega_h}}\right) }} \bigcap_{T\subseteq \mathcal{I}: T^c \cap \mathcal{D} \ne \emptyset } \left\{ R_\mathcal{I}\in \mathbb{R}_+^N\left| \,  \parbox[c]{2.5in}{$\sum\limits_{i\in T}R_i  \le I_{p_{X_{\mathcal{E}}, Y_{\mathcal{E}}}} (X_{T\times \mathcal{I}};Y_{\mathcal{I}\times T^c}|X_{T^c\times \mathcal{I}})$} \right. \right\}\label{Rout}
\end{align}
be the classical cut-set bound for the DM-MMN with independent channels. Then for any channel operation sequence $\boldsymbol{\pi}$, we have
\[
\mathcal{C}^{\boldsymbol{\pi}}\subseteq\mathcal{R}_{\text{out}}.
\]
Hence,
 \[
\mathcal{C}\subseteq\mathcal{R}_{\text{out}}.
\]
\end{Theorem}
\smallskip
\begin{Remark}\label{remark1}
Theorem~\ref{thmCapacityRegionMulticast} subsumes the classical cut-set bound for $\alpha=1$, i.e.,
\begin{equation*}
\mathcal{C}\subseteq  \bigcup_{p_{X_{\mathcal{E}}}  }
\bigcap_{T\subseteq \mathcal{I}: T^c \cap \mathcal{D} \ne \emptyset } \left\{ R_\mathcal{I}\in \mathbb{R}_+^N\left| \,  \parbox[c]{2.7 in}{$\sum\limits_{i\in T}R_i  \le I_{p_{X_{\mathcal{E}}}q^{(1)}_{Y_{\mathcal{E}} | X_{\mathcal{E}}}} (X_{T\times \mathcal{I}};Y_{\mathcal{I}\times T^c}|X_{T^c\times \mathcal{I}})$}\!\! \right. \right\}.
\end{equation*}
\end{Remark}
\begin{Remark}\label{remark2}
Recall that the BSC-IF can be viewed as a multicast network as explained in Section~\ref{introduction}. Using the formulation of BSC-IF in Examples~\ref{example1} and~\ref{example2}, we obtain by Theorem~\ref{thmCapacityRegionMulticast} that $R_1$ is bounded above by the capacity of the BSC and $R_2$ is bounded above by the capacity of the DMC regardless of what order the two channels are operated.
\end{Remark}
\begin{Remark}\label{remark3}
The cut-set bound in Theorem~\ref{thmCapacityRegionMulticast} is tight for the MMN consisting of independent DMCs \cite{networkEquivalencePartI}. This will be shown in Section~\ref{sectionMMNwithIndepDMCs} after the proof of Theorem~\ref{thmCapacityRegionMulticast} is presented in the next section.
\end{Remark}
\begin{Remark}\label{remark4}
The cut-set bound in Theorem~\ref{thmCapacityRegionMulticast} can easily be generalized for (i) any multiple multicast demand where each source multicasts a single message and the destinations want to decode different subsets of the messages; (ii) any multiple unicast demand where each node transmits $N-1$ independent messages to the other $N-1$ nodes and each message is decoded by one node only.
\end{Remark}

%

\section{Proof of Theorem~\ref{thmCapacityRegionMulticast}}\label{sectionInnerOuterBound}
\subsection{Using Fano's Inequality to Bound Sum-Rates}
Fix a channel operation sequence $\boldsymbol{\pi}$. Let $R_{\mathcal{I}}$ be a $\boldsymbol{\pi}$-achievable rate tuple for the DM-MMN denoted by $(\mathcal{X}_\mathcal{E}, \mathcal{Y}_\mathcal{E}, \alpha, \boldsymbol{\Omega}, \boldsymbol q)$. By Definitions~\ref{defAchievableRate} and~\ref{defCapacityRegion}, there exists a sequence of $(\boldsymbol{\pi}, B, n, M_{\mathcal{I}})$-codes such that
\begin{equation}
 \liminf_{n\rightarrow \infty} \frac{\log M_{i}}{n} \ge R_{i} \label{eqnRateLimit}
\end{equation}
for each $i\in\mathcal{I}$ and
\begin{equation}
 \lim_{n\rightarrow \infty} P_{\text{err}}^{n} = 0. \label{eqnProbLimit}
 \end{equation}
 Since $\boldsymbol{\pi}$ is a permutation of $(1, 2, \ldots, \alpha)$ by Definition~\ref{defChannelOperationSequence} and $\cup_{h=1}^\alpha \Omega_h=\mathcal{E}$ by Definition~\ref{defOrderedPartition}, it follows that
 \begin{equation}
 \bigcup_{h=1}^\alpha \Omega_{\pi(h)}=  \bigcup_{h=1}^\alpha \Omega_h=\mathcal{E}. \label{unionOmega=edgeSet}
 \end{equation}
Fix~$n$ and the corresponding $(\boldsymbol{\pi}, B, n, M_{\mathcal{I}})$-code, and let $p_{W_\mathcal{I}, X_\mathcal{E}^n, Y_\mathcal{E}^n, \hat W_{\mathcal{I}\times \mathcal{I}}}$ be the probability distribution induced by the code. Fix any $T\subseteq \mathcal{I}$ such that $T^c\cap \mathcal{D}\ne \emptyset$, and let $d$ denote a node in $T^c\cap \mathcal{D}$.
For the $(\boldsymbol{\pi}, B, n, M_{\mathcal{I}})$-code, since the $N$ messages $W_{1}, W_{2}, \ldots, W_{N}$ are independent, we have
\begin{align}
\sum_{i\in T} \log M_{i}
&= I_{p_{W_\mathcal{I}, Y_{\mathcal{E}}^n}}(W_{T}; Y_{\mathcal{I}\times T^c}^n|W_{T^c}) +
H_{p_{W_\mathcal{I}, Y_{\mathcal{E}}^n}}(W_{T}|Y_{\mathcal{I}\times T^c}^n,W_{T^c}) \notag\\
&\le I_{p_{W_\mathcal{I}, Y_{\mathcal{E}}^n}}(W_{T}; Y_{\mathcal{I}\times T^c}^n|W_{T^c}) + H_{p_{W_\mathcal{I}, Y_{\mathcal{E}}^n}}(W_{T}|Y_{\mathcal{I}\times \{d\}}^n, W_{d})\notag \\
 &\le   I_{p_{W_\mathcal{I}, Y_{\mathcal{E}}^n}}(W_{T};  Y_{\mathcal{I}\times T^c}^n|W_{T^c}) + 1+ P_{\text{err}}^n \sum_{i\in T} \log M_{i},
\label{cutseteqnSet1}
\end{align}
where the last inequality follows from Fano's inequality (cf.\ Definition~\ref{defErrorProbability}).
\subsection{Using Discrete Memoryless Property to Simplify the Upper Bound}
Following \eqref{cutseteqnSet1} and omitting the subscripts for the entropy and mutual information terms, we consider
\begin{align}
 & I(W_T; Y_{\mathcal{I}\times T^c}^n|W_{T^c})\notag \\
&=  \sum_{k=1}^n  (H(Y_{\mathcal{I}\times T^c,k}|W_{T^c}, Y_{\mathcal{I}\times T^c}^{k-1})  - H(Y_{\mathcal{I}\times T^c,k}|W_{\mathcal{I}} ,Y_{\mathcal{I}\times T^c}^{k-1})) \notag \\
& \stackrel{\text{(a)}}{=} \sum_{k=1}^n  (H(Y_{(\mathcal{I}\times T^c) \cap \Omega_{\boldsymbol{\pi}}^{\alpha} ,k}|W_{T^c}, Y_{\mathcal{I}\times T^c}^{k-1}) -H(Y_{(\mathcal{I}\times T^c) \cap \Omega_{\boldsymbol{\pi}}^{\alpha} ,k}|W_{\mathcal{I}} ,Y_{\mathcal{I}\times T^c}^{k-1})) \notag \\
& = \sum_{k=1}^n \sum_{h: \Omega_{\pi(h)}\cap(\mathcal{I}\times T^c)\ne \emptyset} (H(Y_{(\mathcal{I}\times T^c)\cap \Omega_{\pi(h)} ,k}| W_{T^c}, Y_{\mathcal{I}\times T^c}^{k-1}, Y_{(\mathcal{I}\times T^c)\cap \Omega_{\boldsymbol{\pi}}^{h-1} ,k})  \notag\\*
& \hspace{1.7 in}- H(Y_{(\mathcal{I}\times T^c)\cap \Omega_{\pi(h)},k}|W_{\mathcal{I}}, Y_{\mathcal{I}\times T^c}^{k-1}, Y_{(\mathcal{I}\times T^c) \cap \Omega_{\boldsymbol{\pi}}^{h-1} ,k} ))
\label{cutsetstatement2}
\end{align}
where (a) follows from \eqref{defOmegaSuperscript} and \eqref{unionOmega=edgeSet}.
  Following~\eqref{cutsetstatement2}, we consider
\begin{align}
& H(Y_{(\mathcal{I}\times T^c)\cap \Omega_{\pi(h)} ,k}|W_{T^c}, Y_{\mathcal{I}\times T^c}^{k-1}, Y_{(\mathcal{I}\times T^c)\cap \Omega_{\boldsymbol{\pi}}^{h-1} ,k}) \notag \\
& \quad\stackrel{\text{(a)}}{=} H(Y_{(\mathcal{I}\times T^c)\cap \Omega_{\pi(h)} ,k}|W_{T^c}, Y_{\mathcal{I}\times T^c}^{k-1}, Y_{(\mathcal{I}\times T^c)\cap \Omega_{\boldsymbol{\pi}}^{h-1} ,k}, X_{(T^c \times \mathcal{I})\cap \Omega_{\pi(h)},k} )\notag \\
&\quad \le  H(Y_{(\mathcal{I}\times T^c)\cap \Omega_{\pi(h)} ,k}|X_{(T^c \times \mathcal{I})\cap \Omega_{\pi(h)},k})\label{cutsetstatement3}
\end{align}
and
\begin{align}
& H(Y_{(\mathcal{I}\times T^c)\cap \Omega_{\pi(h)} ,k}|W_\mathcal{I}, Y_{\mathcal{I}\times T^c}^{k-1}, Y_{(\mathcal{I}\times T^c)\cap \Omega_{\boldsymbol{\pi}}^{h-1} ,k})\notag \\
&\quad \ge H(Y_{(\mathcal{I}\times T^c)\cap \Omega_{\pi(h)} ,k}|W_\mathcal{I}, Y_{\mathcal{I}\times T^c}^{k-1}, Y_{(\mathcal{I}\times T^c)\cap \Omega_{\boldsymbol{\pi}}^{h-1} ,k}, X_{\Omega_{\pi(h)},k}) \notag \\
& \quad\stackrel{\text{(b)}}{=}\! H(Y_{(\mathcal{I}\times T^c)\cap \Omega_{\pi(h)} ,k}| X_{\Omega_{\pi(h)},k}) \label{cutsetstatement4}
\end{align}
for each $k\in\{1, 2, \ldots, n\}$ and each $h$ that satisfies $\Omega_{\pi(h)}\cap (\mathcal{I}\times T^c)\ne \emptyset$, where
\begin{enumerate}
\item[(a)] follow from the fact by Proposition \ref{propositionXFunctionOfY} that $X_{(T^c \times \mathcal{I})\cap \Omega_{\pi(h)},k}$ is a function of
   $
     (W_{T^c}, Y_{\mathcal{I}\times T^c}^{k-1},  Y_{(\mathcal{I}\times T^c) \cap \Omega_{\boldsymbol{\pi}}^{h-1} ,k})$.
\item[(b)] is due to the following Markov chain implied by \eqref{memorylessStatement} in Definition~\ref{defDiscreteMemoryless}:
\[
\left((W_\mathcal{I}, Y_{\mathcal{I}\times T^c}^{k-1}, Y_{(\mathcal{I}\times T^c)\cap \Omega_{\boldsymbol{\pi}}^{h-1} ,k})  \rightarrow X_{\Omega_{\pi(h)},k} \rightarrow Y_{\Omega_{\pi(h)},k}\right)_p.
\]

\end{enumerate}
Combining \eqref{cutseteqnSet1}, \eqref{cutsetstatement2}, \eqref{cutsetstatement3} and \eqref{cutsetstatement4} and using the fact that $\boldsymbol{\pi}$ is a permutation of $(1, 2, \ldots, \alpha)$, we obtain
\begin{align}
  & (1-P_{\text{err}}^n)\sum_{i\in T}\frac{1}{n}\log M_i
 \notag\\
 &\quad \le \frac{1}{n}\sum_{k=1}^n \sum_{h:\Omega_{h}\cap (\mathcal{I}\times T^c)\ne \emptyset} \big(H(Y_{(\mathcal{I}\times T^c)\cap \Omega_{h} ,k}| X_{(T^c \times \mathcal{I})\cap \Omega_{h},k}) - H(Y_{(\mathcal{I}\times T^c)\cap \Omega_{h} ,k}| X_{\Omega_{h},k})\big)  \label{cutsetStatement5*}\\
 & \quad \le \frac{1}{n}\sum_{k=1}^n \sum_{h=1}^\alpha \big(H(Y_{(\mathcal{I}\times T^c)\cap \Omega_{h} ,k}| X_{(T^c \times \mathcal{I})\cap \Omega_{h},k}) - H(Y_{(\mathcal{I}\times T^c)\cap \Omega_{h} ,k}| X_{\Omega_{h},k})\big).  \label{cutsetStatement5}
\end{align}
%
%
\subsection{Introducing Time-Sharing Distribution to Single-Letterize Entropy Terms}
Recalling that $p_{W_\mathcal{I}, X_\mathcal{E}^n, Y_\mathcal{E}^n, \hat W_{\mathcal{I}\times \mathcal{I}}}$ is the probability distribution induced by the code, we define the time-sharing distribution
\begin{equation}
 p_{Q_n}(k) \triangleq \frac{1}{n} \label{defDistQ}
\end{equation}
for all $k\in\{1, 2, \ldots, n\}$.
In addition, we define
\begin{equation}
s_{Q_n, X_{\mathcal{E}, Q_n}, Y_{\mathcal{E},Q_n}}(k, x_{\mathcal{E}}, y_{\mathcal{E}}) \triangleq  p_{Q_n}(k) p_{X_{\mathcal{E},k}, Y_{\mathcal{E},k}}(x_{\mathcal{E}}, y_{\mathcal{E}}) \label{defDistS}
\end{equation}
for all $k\in\{1, 2, \ldots, n\}$, $x_{\mathcal{E}}\in \mathcal{X}_{\mathcal{E}}$ and $y_{\mathcal{E}}\in \mathcal{Y}_{\mathcal{E}}$.
Following \eqref{cutsetStatement5}, we consider the following chain of inequalities for each $h\in\{1, 2, \ldots, \alpha\}$:
\begin{align}
& \frac{1}{n}\sum_{k=1}^n (H_{p_{X_\mathcal{E}^n, Y_\mathcal{E}^n}}(Y_{(\mathcal{I}\times T^c)\cap \Omega_{h} ,k}| X_{(T^c \times \mathcal{I})\cap \Omega_{h},k}) - H_{p_{X_\mathcal{E}^n, Y_\mathcal{E}^n}}(Y_{(\mathcal{I}\times T^c)\cap \Omega_{h} ,k}| X_{\Omega_{h},k})) \notag\\
& \quad \stackrel{\text{(a)}}{=}\frac{1}{n}\sum_{k=1}^n (H_{s_{Q_n, X_{\mathcal{E},Q_n}, Y_{\mathcal{E},Q_n}}}(Y_{(\mathcal{I}\times T^c)\cap \Omega_{h} ,Q_n}| X_{(T^c \times \mathcal{I})\cap \Omega_{h},Q_n}, Q_n=k) \notag\\
 &\qquad \qquad \qquad  \qquad- H_{s_{Q_n, X_{\mathcal{E},Q_n}, Y_{\mathcal{E},Q_n}}}(Y_{(\mathcal{I}\times T^c)\cap \Omega_{h} ,Q_n}| X_{\Omega_{h},Q_n}, Q_n=k)) \notag\\
 & \quad \stackrel{\eqref{defDistQ}}{=} H_{s_{Q_n, X_{\mathcal{E},Q_n}, Y_{\mathcal{E},Q_n}}}(Y_{(\mathcal{I}\times T^c)\cap \Omega_{h} ,Q_n}| X_{(T^c \times \mathcal{I})\cap \Omega_{h},Q_n}, Q_n) \notag\\
 &\qquad \qquad \qquad - H_{s_{Q_n, X_{\mathcal{E},Q_n}, Y_{\mathcal{E},Q_n}}}(Y_{(\mathcal{I}\times T^c)\cap \Omega_{h} ,Q_n}| X_{\Omega_{h},Q_n}, Q_n) \notag\\
 &\quad \le H_{s_{Q_n, X_{\mathcal{E},Q_n}, Y_{\mathcal{E},Q_n}}}(Y_{(\mathcal{I}\times T^c)\cap \Omega_{h} ,Q_n}| X_{(T^c \times \mathcal{I})\cap \Omega_{h},Q_n}) \notag\\
 &\qquad \qquad \qquad - H_{s_{Q_n, X_{\mathcal{E},Q_n}, Y_{\mathcal{E},Q_n}}}(Y_{(\mathcal{I}\times T^c)\cap \Omega_{h} ,Q_n}| X_{\Omega_{h},Q_n}, Q_n) \notag\\
 &\quad \stackrel{\text{(b)}}{=} H_{s_{Q_n, X_{\mathcal{E},Q_n}, Y_{\mathcal{E},Q_n}}}(Y_{(\mathcal{I}\times T^c)\cap \Omega_{h} ,Q_n}| X_{(T^c \times \mathcal{I})\cap \Omega_{h},Q_n}) \notag\\
 &\qquad \qquad \qquad - H_{s_{Q_n, X_{\mathcal{E},Q_n}, Y_{\mathcal{E},Q_n}}}(Y_{(\mathcal{I}\times T^c)\cap \Omega_{h} ,Q_n}| X_{\Omega_{h},Q_n}) \notag\\
 & \quad = I_{s_{X_{\Omega_{h}, Q_n}, Y_{\Omega_{h}, Q_n}}}(X_{(T \times \mathcal{I})\cap \Omega_{h}, Q_n};Y_{(\mathcal{I}\times T^c)\cap \Omega_{h}, Q_n }|X_{(T^c \times \mathcal{I})\cap \Omega_{h},Q_n}) \label{cutsetStatement6}
\end{align}
where
\begin{enumerate}
\item[(a)] is due to the following fact by \eqref{defDistS}:
\[
s_{X_{\mathcal{E}, Q_n}, Y_{\mathcal{E},Q_n}|Q_n}(x_{\mathcal{E}}, y_{\mathcal{E}}|k) = p_{X_{\mathcal{E},k}, Y_{\mathcal{E},k}}(x_{\mathcal{E}}, y_{\mathcal{E}})
\]
for each $k$, $x_{\mathcal{E}}$ and $y_{\mathcal{E}}$.
\item[(b)] is due to the following Markov chain implied by~\eqref{defDistS}:
\[
\left(Q_n \rightarrow X_{\Omega_{h}, Q_n} \rightarrow  Y_{(\mathcal{I}\times T^c)\cap \Omega_{h} ,Q_n}\right)_s.
\]
\end{enumerate}
Combining \eqref{cutsetStatement5} and \eqref{cutsetStatement6}, we obtain
\begin{equation}
(1-P_{\text{err}}^n)\sum_{i\in T}\frac{1}{n}\log M_i \le \sum_{h=1}^\alpha I_{s_{X_{\Omega_h, Q_n}, Y_{\Omega_h, Q_n}}}(X_{(T \times \mathcal{I})\cap \Omega_h, Q_n};Y_{(\mathcal{I}\times T^c)\cap \Omega_h, Q_n }|X_{(T^c \times \mathcal{I})\cap \Omega_h,Q_n}) \label{cutsetStatement7}
\end{equation}
where for each $h\in\{1, 2, \ldots, \alpha\}$, $s_{X_{\Omega_h, Q_n}, Y_{\Omega_h, Q_n}}$ is a distribution on $(\mathcal{X}_{\Omega_h}, \mathcal{Y}_{\Omega_h})$ that satisfies
\begin{align}
& s_{X_{\Omega_h, Q_n}, Y_{\Omega_h, Q_n}}(x_{\Omega_h}, y_{\Omega_h}) \notag\\
& \stackrel{\eqref{defDistS}}{=} \frac{1}{n}\sum_{k=1}^n p_{X_{\Omega_h,k}, Y_{\Omega_h,k}}(x_{\Omega_h}, y_{\Omega_h}) \notag\\
& \stackrel{\eqref{memorylessStatement}}{=} \frac{1}{n}\sum_{k=1}^n \left(p_{X_{\Omega_h,k}}(x_{\Omega_h})q_{Y_{\Omega_h}| X_{\Omega_h}}^{(h)}( y_{\Omega_h}|x_{\Omega_h})\right) \notag\\
  & = \left(\frac{1}{n}\sum_{k=1}^n p_{X_{\Omega_h,k}}(x_{\Omega_h})\right)q_{Y_{\Omega_h}| X_{\Omega_h}}^{(h)}( y_{\Omega_h}|x_{\Omega_h}). \label{eqnDistSinTermsOfP}
\end{align}
Let $\{n_{\ell}\}_{\ell=1}^\infty$ be a subsequence of $\{n\}_{n=1}^{\infty}$ such that $s_{X_{\Omega_{h}, Q_{n_\ell}}, Y_{\Omega_{h}, Q_{n_\ell}}}$ converges with respect to the $\mathcal{L}_1$-distance for all $h\in\{1, 2, \ldots, \alpha\}$, and we define the limit of $s_{X_{\Omega_{h}, Q_{n_\ell}}, Y_{\Omega_{h}, Q_{n_\ell}}}$ as
\begin{equation}
\bar s_{X_{\Omega_h}, Y_{\Omega_h}}(x_{\Omega_h}, y_{\Omega_h})\triangleq \lim_{\ell\rightarrow \infty} s_{X_{\Omega_{h}, Q_{n_\ell}}, Y_{\Omega_{h}, Q_{n_\ell}}}(x_{\Omega_h}, y_{\Omega_h}) \label{defBarDistS}
\end{equation}
for all $(x_{\Omega_h}, y_{\Omega_h})\in \mathcal{X}_{\Omega_h}\times \mathcal{Y}_{\Omega_h}$ and all $h\in\{1, 2, \ldots, \alpha\}$. Since \[I_{s_{X_{\Omega_h}, Y_{\Omega_h}}}(X_{(T \times \mathcal{I})\cap \Omega_{h}};Y_{(\mathcal{I}\times T^c)\cap \Omega_{h}}|X_{(T^c \times \mathcal{I})\cap \Omega_{h}})\] is a continuous functional of $s_{X_{\Omega_h}, Y_{\Omega_h}}$ for each $h\in\{1, 2, \ldots, \alpha\}$, we obtain from \eqref{cutsetStatement7}, \eqref{defBarDistS}, \eqref{eqnRateLimit} and \eqref{eqnProbLimit} that
\begin{align}
\sum_{i\in T}R_i \le \sum_{h=1}^\alpha I_{\bar s_{X_{\Omega_h}, Y_{\Omega_h}}}(X_{(T \times \mathcal{I})\cap \Omega_h};Y_{(\mathcal{I}\times T^c)\cap \Omega_h}|X_{(T^c \times \mathcal{I})\cap \Omega_h}).  \label{st2InTheorem}
\end{align}
Since
\[
 \bar s_{X_{\Omega_h}, Y_{\Omega_h}}=\bar s_{X_{\Omega_{h}}} q^{(h)}_{Y_{\Omega_{h}} | X_{\Omega_{h}}}
\]
for all $h\in\{1, 2, \ldots,\alpha\}$ by \eqref{eqnDistSinTermsOfP} and \eqref{defBarDistS}, it follows from \eqref{st2InTheorem} that
\begin{align}
\sum_{i\in T} R_i & \le\sum_{h=1}^\alpha  I_{\bar s_{X_{\Omega_{h}}} q^{(h)}_{Y_{\Omega_{h}} | X_{\Omega_{h}}} }(X_{(T \times \mathcal{I})\cap \Omega_{h}}; Y_{(\mathcal{I}\times T^c)\cap \Omega_{h} }|X_{(T^c \times \mathcal{I})\cap \Omega_{h}}) . \label{st4InTheorem}
\end{align}
Define
\begin{align}
\hat p_{X_{\mathcal{E}}, Y_{\mathcal{E}}}&  \triangleq
\prod_{h=1}^\alpha \left(\bar s_{X_{\Omega_h}}
q^{(h)}_{Y_{\Omega_h} | X_{\Omega_h}}\right)   \label{defPHat}\\
& \stackrel{\eqref{defPHat} }{=} \prod_{h=1}^\alpha \left(\hat p_{X_{\Omega_h}}
q^{(h)}_{Y_{\Omega_h} | X_{\Omega_h}}\right).   \label{defPHat*}
\end{align}
It then follows from \eqref{st4InTheorem} that
\begin{align}
\sum_{i\in T} R_i \le \sum_{h=1}^\alpha  I_{\hat p_{X_{\mathcal{E}}, Y_{\mathcal{E}}}}\!\!(X_{(T \times \mathcal{I})\cap \Omega_h};Y_{(\mathcal{I}\times T^c)\cap \Omega_h }| X_{(T^c \times \mathcal{I})\cap \Omega_h}). \label{st5InTheorem}
\end{align}
%
In order to simplify \eqref{st5InTheorem}, we consider
\begin{align}
&  I_{\hat p_{X_{\mathcal{E}}, Y_{\mathcal{E}}}} (X_{T\times \mathcal{I}};Y_{\mathcal{I}\times T^c}|X_{T^c\times \mathcal{I}})\notag\\
& \stackrel{\eqref{unionOmega=edgeSet}}{=} \sum_{h=1}^\alpha I_{\hat p_{X_{\mathcal{E}}, Y_{\mathcal{E}}}} (X_{T\times \mathcal{I}};Y_{(\mathcal{I}\times T^c)\cap \Omega_h}|X_{T^c\times \mathcal{I}}, Y_{(\mathcal{I}\times T^c)\cap \Omega^{h-1}})\notag\\
& = \sum_{h=1}^\alpha (H_{\hat p_{X_{\mathcal{E}}, Y_{\mathcal{E}}}} (Y_{(\mathcal{I}\times T^c)\cap \Omega_h}|X_{T^c\times \mathcal{I}}, Y_{(\mathcal{I}\times T^c)\cap \Omega^{h-1}})- H_{\hat p_{X_{\mathcal{E}}, Y_{\mathcal{E}}}} (Y_{(\mathcal{I}\times T^c)\cap \Omega_h}|X_{\mathcal{E}}, Y_{(\mathcal{I}\times T^c)\cap \Omega^{h-1}})) \notag\\
& \stackrel{\text{(a)}}{=} \sum_{h=1}^\alpha (H_{\hat p_{X_{\mathcal{E}}, Y_{\mathcal{E}}}} (Y_{(\mathcal{I}\times T^c)\cap \Omega_h}|X_{(T^c\times \mathcal{I})\cap \Omega_h}) - H_{\hat p_{X_{\mathcal{E}}, Y_{\mathcal{E}}}} (Y_{(\mathcal{I}\times T^c)\cap \Omega_h}|X_{\mathcal{E}\cap \Omega_h})) \notag\\
&=\sum_{h=1}^\alpha I_{\hat p_{X_{\mathcal{E}}, Y_{\mathcal{E}}}}(X_{(T \times \mathcal{I})\cap \Omega_h};Y_{(\mathcal{I}\times T^c)\cap \Omega_h }| X_{(T^c \times \mathcal{I})\cap \Omega_h}) \label{st6InTheorem}
\end{align}
where
(a) follows from the fact by \eqref{defPHat*} that $(X_{\Omega_\ell}, Y_{\Omega_\ell})$ and $(X_{\Omega_m}, Y_{\Omega_m})$ are independent for all $\ell\ne m$ under the distribution $\hat p_{X_{\mathcal{E}}, Y_{\mathcal{E}}}$.
Combining \eqref{st5InTheorem} and \eqref{st6InTheorem}, we obtain
\begin{equation}
\sum_{i\in T} R_i \le  I_{\hat p_{X_{\mathcal{E}}, Y_{\mathcal{E}}}} (X_{T\times \mathcal{I}};Y_{\mathcal{I}\times T^c}|X_{T^c\times \mathcal{I}}). \label{st7InTheorem}
\end{equation}
Since $\hat p_{X_{\mathcal{E}}, Y_{\mathcal{E}}}$ satisfies \eqref{defPHat*} and depends on only the sequence of $(\boldsymbol{\pi},B, n, M_{\mathcal{I}})$-codes but not on $T$, \eqref{st7InTheorem} holds for all $T\subseteq \mathcal{I}$ that satisfies $T^c\cap \mathcal{D}\ne \emptyset$, which implies from \eqref{defPHat*} and \eqref{Rout} that $R_\mathcal{I}\in \mathcal{R}_{\text{out}}$. This completes the proof.

\section{Multicast Network Consisting of Independent DMCs} \label{sectionMMNwithIndepDMCs}
The MMN consisting of independent DMCs \cite{networkEquivalencePartI} consists of $N^2$ channels where each channel is associated with a directed edge. Define \begin{equation}
\Omega_{N(i-1)+j}\triangleq\{(i,j)\} \label{defOmegaij}
 \end{equation}
 for each $(i,j)\in \mathcal{I}\times \mathcal{I}$ such that $q_{Y_{\Omega_{N(i-1)+j}}|X_{\Omega_{N(i-1)+j}}}^{(N(i-1)+j)}$ characterizes the DMC associated with edge~$(i,j)$. Then, this network can be viewed as a discrete network $(\mathcal{X}_\mathcal{E}, \mathcal{Y}_\mathcal{E},N^2, \boldsymbol{\Omega}, \boldsymbol{q})$ according to Definition~\ref{defDiscreteNetwork}, where $  \boldsymbol{\Omega}$ and $\boldsymbol{q}$ are defined as
  \[
   \boldsymbol{\Omega} \triangleq (\Omega_1, \Omega_2, \ldots, \Omega_{N^2})
  \]
  and
  \[
   \boldsymbol{q} \triangleq (q^{(1)}, q^{(2)}, \ldots, q^{(N^2)})
  \]
  respectively. We assume that this network satisfies the discrete memoryless property stated in Definition~\ref{defDiscreteMemoryless}.
  Let $\mathcal{C}$ denote the capacity region of this network. To simplify notation, we let
\begin{equation}
q_{Y_{(i,j)}|X_{(i,j)}}\triangleq  q_{Y_{\Omega_{N(i-1)+j}}|X_{\Omega_{N(i-1)+j}}}^{(N(i-1)+j)} \label{defChannelMMNindDMCs}
\end{equation}
and let
\begin{equation}
C_{(i,j)}\triangleq \max_{p_{X_{(i,j)}}}I_{p_{X_{(i,j)}}q_{Y_{(i,j)}|X_{(i,j)}} }(X_{(i,j)};Y_{(i,j)}) \label{defCij}
\end{equation}
be the capacity of channel $q_{Y_{(i,j)}|X_{(i,j)}}$ for each $(i,j)\in  \mathcal{I}\times \mathcal{I}$.
\smallskip
Combining existing results with Theorem~\ref{thmCapacityRegionMulticast}, we fully characterize the capacity region of the MMN with independent DMCs and zero-delay edges in the following theorem.
\smallskip
\begin{Theorem} \label{thmMMNwithIndDMCs}
For the MMN consisting of independent DMCs with zero-delay edges under multicast demand $(\mathcal{V}, \mathcal{D})$, we have
\begin{equation*}
\mathcal{C} = \mathcal{C}^{\mathbf{1}}=
\bigcap_{T\subseteq \mathcal{I}: T^c \cap \mathcal{D} \ne \emptyset } \left\{ R_\mathcal{I}\in \mathbb{R}_+^N\left| \,  \parbox[c]{1.6 in}{$\sum\limits_{i\in T}R_i  \le \sum\limits_{(i,j)\in T\times T^c} C_{(i,j)}$} \right. \right\},
\end{equation*}
where $\mathcal{C}$ and $\mathcal{C}^{\mathbf{1}}$ denote the capacity region and the $\mathbf{1}$-capacity region respectively.
\end{Theorem}
\begin{Remark}
It was shown by Effros \cite{EffrosIndependentDelay} that under the positive-delay assumption in the classical setting, the set of achievable rate tuples for the MMN with independent DMCs does not depend on the amount of positive delay incurred by each edge on each other edge. Therefore, Theorem~\ref{thmMMNwithIndDMCs} together with Effros's result implies that the set of achievable rate tuples for the MMN with independent DMCs does not depend on the amount of delay incurred by each edge on each other edge even when zero-delay edges are present.
\end{Remark}

The achievability and converse proofs of Theorem~\ref{thmMMNwithIndDMCs} are presented in the next two subsections respectively.
\subsection{Achievability} \label{subsectionAchIndDMCs}
 Define
 \begin{equation}
 \mathcal{R}^{\mathbf{1}}\triangleq
\bigcap_{T\subseteq \mathcal{I}: T^c \cap \mathcal{D} \ne \emptyset } \left\{ R_\mathcal{I}\in \mathbb{R}_+^N\left| \,  \parbox[c]{1.75 in}{$\sum_{i\in T}R_i  \le \sum\limits_{(i,j)\in T\times T^c} C_{(i,j)}$} \right. \right\}. \label{defR1ach}
  \end{equation}
Our goal is to prove
   \begin{equation}
 \mathcal{C}\supseteq \mathcal{C}^{\mathbf{1}} \supseteq \mathcal{R}^{\mathbf{1}}. \label{CsupersetR1}
 \end{equation}
The proof combines existing results from network equivalence theory in \cite{networkEquivalencePartI} and noisy network coding (NNC) in \cite{YAG11,Lim11}. Consider a deterministic counterpart of the MMN consisting of independent DMCs by replacing every DMC $q_{Y_{(i,j)}|X_{(i,j)}}$ in the MMN with a noiseless bit pipe whose capacity is equal to $C_{(i,j)}$ (cf.\ \eqref{defOmegaij}, \eqref{defChannelMMNindDMCs} and \eqref{defCij}), and let $\mathcal{C}_{\text{det}}^{\mathbf{1}}$ be the capacity region of the deterministic counterpart network \textit{with unit-delay edges} where $\mathbf{1}$ denotes the all-one delay profile. Letting $\mathcal{C}^{\mathbf{1}}$ denote the $\mathbf{1}$-capacity region of the MMN consisting of independent DMCs, we conclude by using the network equivalence theory \cite{networkEquivalencePartI} that $\mathcal{C}^{\mathbf{1}}=\mathcal{C}_{\text{det}}^{\mathbf{1}}$, which then implies from Definition~\ref{defCapacityRegion} that
 \begin{equation}
 \mathcal{C}\supseteq \mathcal{C}^{\mathbf{1}} = \mathcal{C}_{\text{det}}^{\mathbf{1}}. \label{CsupersetCdet}
 \end{equation}
%
By viewing~$\mathcal{R}^{\mathbf{1}}$ as the cut-set bound for the deterministic counterpart network, it follows from the NNC inner bound in \cite[Sec.~II-A]{Lim11} that $\mathcal{C}_{\text{det}}^{\mathbf{1}}\supseteq\mathcal{R}^{\mathbf{1}}$, which then implies from \eqref{CsupersetCdet} that \eqref{CsupersetR1} holds.

\subsection{Converse}
We will prove
 \begin{equation}
 \mathcal{C}\subseteq  \mathcal{R}^{\mathbf{1}} \label{converseProofForMMNwithDMCs}
 \end{equation}
  by using Theorem~\ref{thmCapacityRegionMulticast}. Suppose $R_\mathcal{I}$ is achievable. It then follows from Theorem~\ref{thmCapacityRegionMulticast}, \eqref{defOmegaij} and \eqref{defChannelMMNindDMCs} that there exists a distribution
\begin{equation}
p_{X_{\mathcal{E}}, Y_{\mathcal{E}}} =
\prod\limits_{i=1}^N \prod\limits_{j=1}^N \left(p_{X_{(i,j)}} \label{eqnDistInConverse}
q_{Y_{(i,j)} | X_{(i,j)}}\right)
\end{equation}
such that for all $T\subseteq{\mathcal{I}}$ that satisfies $T^c\cap \mathcal{D}\ne \emptyset$, we have
\begin{equation}
\sum\limits_{i\in T}R_i  \le I_{p_{X_{\mathcal{E}}, Y_{\mathcal{E}}}} (X_{T\times \mathcal{I}};Y_{\mathcal{I}\times T^c}|X_{T^c\times \mathcal{I}}). \label{eqnCutsetInConverse}
\end{equation}
Consider the following chain of inequalities for each $T\subseteq{\mathcal{I}}$:
\begin{align}
& I_{p_{X_{\mathcal{E}}, Y_{\mathcal{E}}}} (X_{T\times \mathcal{I}};Y_{\mathcal{I}\times T^c}|X_{T^c\times \mathcal{I}}) \notag\\
& = \sum_{i=1}^N I_{p_{X_{\mathcal{E}}, Y_{\mathcal{E}}}} (X_{T\times \mathcal{I}};Y_{\{i\}\times T^c}|X_{T^c\times \mathcal{I}}, Y_{\{i^\prime \in \mathcal{I} | i^\prime < i\}\times T^c}) \notag\\
& = \sum_{i=1}^N \sum_{j\in T^c} I_{p_{X_{\mathcal{E}}, Y_{\mathcal{E}}}} (X_{T\times \mathcal{I}};Y_{(i, j)}|X_{T^c\times \mathcal{I}}, Y_{\{i^\prime \in \mathcal{I} | i^\prime < i\}\times T^c}, Y_{\{i\}\times \{j^\prime \in T^c|j^\prime <j\}}) \notag\\
& \le \sum_{i=1}^N \sum_{j\in T^c} \left(H_{p_{X_{\mathcal{E}}, Y_{\mathcal{E}}}}(Y_{(i, j)}|X_{T^c\times \mathcal{I}})- H_{p_{X_{\mathcal{E}}, Y_{\mathcal{E}}}}(Y_{(i, j)}|X_\mathcal{E}, Y_{\{i^\prime \in \mathcal{I} | i^\prime < i\}\times T^c}, Y_{\{i\}\times \{j^\prime \in T^c|j^\prime <j\}})\right)\notag\\
&\stackrel{\text{(a)}}{=}  \sum_{i=1}^N \sum_{j\in T^c} \left(H_{p_{X_{\mathcal{E}}, Y_{\mathcal{E}}}}(Y_{(i, j)}|X_{T^c\times \mathcal{I}})- H_{p_{X_{\mathcal{E}}, Y_{\mathcal{E}}}}(Y_{(i, j)}|X_{(i,j)})\right)\notag\\
&\stackrel{\text{(b)}}{=}  \sum_{i\in T} \sum_{j\in T^c} \left(H_{p_{X_{\mathcal{E}}, Y_{\mathcal{E}}}}(Y_{(i, j)}|X_{T^c\times \mathcal{I}})- H_{p_{X_{\mathcal{E}}, Y_{\mathcal{E}}}}(Y_{(i, j)}|X_{(i,j)})\right)\notag\\
&\le \sum_{i\in T} \sum_{j\in T^c} \left(H_{p_{X_{\mathcal{E}}, Y_{\mathcal{E}}}}(Y_{(i, j)})- H_{p_{X_{\mathcal{E}}, Y_{\mathcal{E}}}}(Y_{(i, j)}|X_{(i,j)})\right)\notag\\
& = \sum_{i\in T} \sum_{j\in T^c}I_{p_{X_{(i,j)}, Y_{(i,j)}}}(X_{(i,j)};Y_{(i,j)})\notag\\
&\stackrel{\eqref{eqnDistInConverse}}{=}  \sum_{i\in T} \sum_{j\in T^c}I_{p_{X_{(i,j)}}q_{Y_{(i,j)}|X_{(i,j)}}}(X_{(i,j)};Y_{(i,j)}) \notag\\
&\stackrel{\eqref{defCij}}{\le}  \sum_{(i,j)\in T\times T^c}C_{(i,j)},
\label{eqn1InConverse}
\end{align}
where
\begin{enumerate}
\item[(a)] follows from the fact by \eqref{eqnDistInConverse} that
\begin{equation*}
\left((X_{\mathcal{E}\setminus \{(i,j)\}}, Y_{\mathcal{E}\setminus \{(i,j)\}}) \rightarrow X_{(i,j)}\rightarrow Y_{(i,j)} \right)_p
\end{equation*}
forms a Markov chain for all $(i,j)\in \mathcal{E}$.
\item[(b)] follows from the fact that for all $(i,j)\in T^c\times T^c$,
\begin{align*}
 H_{p_{X_{\mathcal{E}}, Y_{\mathcal{E}}}}(Y_{(i, j)}|X_{T^c\times \mathcal{I}})
\le H_{p_{X_{\mathcal{E}}, Y_{\mathcal{E}}}}(Y_{(i, j)}|X_{(i,j)}).
\end{align*}
\end{enumerate}
Combining \eqref{eqnCutsetInConverse}, \eqref{eqn1InConverse} and \eqref{defR1ach}, we have $R_\mathcal{I}\in \mathcal{R}^{\mathbf{1}}$. This completes the proof of \eqref{converseProofForMMNwithDMCs}.

\section{Multicast Network Consisting of Independent AWGN Channels} \label{sectionGaussianChannels}
In this section, we generalize the result in the previous section to multicast networks consisting of independent additive white Gaussian noise (AWGN) channels with zero-delay edges. The \emph{MMN consisting of independent AWGN channels} consists of $N^2$ channels, where the channel associated with edge $(i,j)$ is an AWGN channel whose noise variance is denoted by~$\sigma_{(i,j)}^2$ for each $(i,j)\in \mathcal{E}$. In each time slot~$k$, node~$i$ transmits $X_{(i,j),k}\in \mathbb{R}$ on edge~$(i,j)$ and receives $Y_{(\ell,i),k}\in \mathbb{R}$ from edge~$(\ell, i)$. In addition, we assume that the channel outputs are independent given their inputs, i.e., the channel noises are independent. Each codeword transmitted on $(i,j)$ is subject to the power constraint
\begin{equation}
\Pr\left\{\frac{1}{n}\sum_{k=1}^n X_{(i,j),k}^2 \le P_{(i,j)}\right\}=1 \label{ineqnPij}
\end{equation}
for each $(i,j)\in \mathcal{E}$,
where $P_{(i,j)}$ denotes the average power available to edge~$(i,j)$. Similarly, each node~$i$ is subject to the power constraint
\begin{equation}
\Pr\left\{\frac{1}{n}\sum_{j=1}^N\sum_{k=1}^n X_{(i,j),k}^2 \le n P_i\right\}=1 \label{ineqnPi}
\end{equation}
for each $i\in \mathcal{I}$, where $P_i$ denotes the average power available to node~$i$. The network is subject to the power constraint
\begin{equation}
\Pr\left\{\frac{1}{n}\sum_{i=1}^N\sum_{j=1}^N\sum_{k=1}^n X_{(i,j),k}^2 \le n P\right\}=1, \label{ineqnP}
\end{equation}
where $P$ denotes the total average power available for transmissions in the network. To facilitate discussion, we write
$
P_{\mathcal{E}}\triangleq (P_{(i,j)}:(i,j)\in \mathcal{E})
$
and
$
P_{\mathcal{I}}\triangleq (P_i : i\in \mathcal{I})$. 
\subsection{Network Model} \label{sectionGaussianModel}
\begin{Definition} \label{defMMNwithAWGN}
The MMN with AWGN channels, denoted by a positive-valued tuple $(\sigma_{(i,j)}^2:(i,j)\in \mathcal{E})$, consists of~$N^2$ AWGN channels denoted by $\left\{q_{Y_{(i,j)}|X_{(i,j)}} : (i,j)\in \mathcal{I}\times \mathcal{I}\right\}$, where $q_{Y_{(i,j)}|X_{(i,j)}}$ characterizes the AWGN channel associated with the directed edge~$(i,j)$ such that
\begin{equation}
q_{Y_{(i,j)}|X_{(i,j)}}(y_{(i,j)}|x_{(i,j)})=\mathcal{N}(y_{(i,j)}-x_{(i,j)}; 0, \sigma_{(i,j)}^2) \label{defChannelMMNindAWGNs}
\end{equation}
for all $(x_{(i,j)}, y_{(i,j)})\in \mathbb{R}^2$.
\end{Definition}
\smallskip

In Definition~\ref{defMMNwithAWGN}, we have not introduced the independence property among the channels yet. We will define the independence among the AWGN channels after we have specified codes that use the network multiple times. To facilitate discussion, we define
\begin{equation}
\Omega_{(i-1)N+j}\triangleq \{(i,j)\} \label{defOmegaijAWGN}
\end{equation}
for each $(i,j)\in\mathcal{I}\times \mathcal{I}$ such that
$
\bigcup_{\ell=1}^{N^2} \Omega_{\ell}=\mathcal{E}
$
and
$
\Omega_{\ell} \cap \Omega_{\ell^\prime}=\emptyset
$
for all $\ell\ne \ell^\prime$. In other words, $(\Omega_1,\Omega_2, \ldots, \Omega_{N^2})$ is an~$N^2$-partition of $\mathcal{E}$.
%
%
As in Definition~\ref{defChannelOperationSequence}, we define a channel operation sequence $\boldsymbol{\pi}$ to be a permutation of $(1,2, \ldots, N^2)$, and define $\Pi$ to be the set of the permutations of $(1,2, \ldots, N^2)$.
For a channel operation sequence~$\boldsymbol{\pi}$ and each delay profile $B$ that is feasible with respect to $\boldsymbol{\pi}$ (cf.\ Definition~\ref{defFeasible}), we define a~$(\boldsymbol{\pi}, B, n, M_\mathcal{I}, P_\mathcal{E}, P_\mathcal{I}, P)$-code similar to Definition~\ref{defCode} with the additional power constraints~\eqref{ineqnPij}, \eqref{ineqnPi} and~\eqref{ineqnP}.
We are ready to formally define the MMN with \textit{independent} AWGN channels.
\smallskip
\begin{Definition}
A MMN with AWGN channels $(\sigma_{(i,j)}^2:(i,j)\in \mathcal{E})$
 is called a \textit{MMN with independent AWGN channels} if the following holds for any
 channel operation sequence $\boldsymbol{\pi}=(\pi(1), \pi(2), \ldots, \pi(N^2))$ and any $(\boldsymbol{\pi}, B, n, M_\mathcal{I}, P_\mathcal{E}, P_\mathcal{I}, P)$-code:

Let $U^{k-1}\triangleq (W_{\mathcal{I}}, X_{\mathcal{E}}^{k-1}, Y_{\mathcal{E}}^{k-1})$ be the collection of random variables that are generated before the $k^{\text{th}}$ time slot. To simplify notation, let
\begin{equation}
\Omega_{\boldsymbol{\pi}}^h \triangleq \bigcup_{m=1}^h \Omega_{\pi(m)}, \label{defOmegaSuperscriptAWGN}
\end{equation}
and let $(i_h,j_h)$ be the unique edge such that
\begin{equation}
\Omega_{\pi(h)}=\{(i_h,j_h)\}.\label{uniqueEdge}
 \end{equation}
 Then, for each $k\in\{1, 2, \ldots, n\}$ and each $h\in \{1, 2, \ldots, \alpha\}$, we have
\begin{align}
 p_{U^{k-1}, X_{\Omega_{\boldsymbol{\pi}}^h ,k}, Y_{\Omega_{\boldsymbol{\pi}}^h ,k}} & =  p_{U^{k-1}, X_{\Omega_{\boldsymbol{\pi}}^h ,k},  Y_{\Omega_{\boldsymbol{\pi}}^{h-1} ,k}}
  p_{Y_{\Omega_{\pi(h)},k}|X_{\Omega_{\pi(h)},k}}, \label{memorylessStatementAWGN} \\*
  & \stackrel{\eqref{uniqueEdge}}{=} p_{U^{k-1}, X_{\Omega_{\boldsymbol{\pi}}^h ,k},  Y_{\Omega_{\boldsymbol{\pi}}^{h-1} ,k}} p_{Y_{(i_h,j_h),k}|X_{(i_h,j_h),k}}
\end{align}
where
\begin{equation}
 p_{Y_{(i_h,j_h),k}|X_{(i_h,j_h),k}}(y_{(i_h,j_h),k}|x_{(i_h,j_h),k})= q_{Y_{(i_h,j_h)}|X_{(i_h,j_h)}}(y_{(i_h,j_h),k}|x_{(i_h,j_h),k}) \label{memorylessStatementAWGN*}
\end{equation}
for all $(x_{(i_h,j_h),k}, y_{(i_h,j_h),k})\in \mathbb{R}^2$ (channel $q_{Y_{(i_h,j_h)}|X_{(i_h,j_h)}}$ was defined in~\eqref{defChannelMMNindAWGNs}).
\end{Definition}
\smallskip

Given a multicast demand $(\mathcal{V}, \mathcal{D})$, we define the \textit{$(\boldsymbol{\pi},B)$-achievability}, the \textit{$\boldsymbol{\pi}$-achievability} and the \text{achievability} of a rate tuple $R_\mathcal{I}$ as in Definition~\ref{defAchievableRate}.  Then, we define the \textit{$(\boldsymbol{\pi},B)$-capacity region} denoted by $\mathcal{C}_{B}^{\boldsymbol{\pi}}$, the \textit{$\boldsymbol{\pi}$-capacity region} denoted by $\mathcal{C}^{\boldsymbol{\pi}}$ and the capacity region
\begin{equation}
\mathcal{C}\triangleq \bigcup_{\boldsymbol{\pi}\in \Pi}\mathcal{C}^{\boldsymbol{\pi}} \label{defCapacityRegionAWGN}
 \end{equation}
 as in Definition~\ref{defCapacityRegion}. The following theorem fully characterizes the capacity region of the MMN with independent AWGNs and zero-delay edges. The achievability and converse proofs of the theorem will be provided in the following two subsections respectively.
\smallskip
\begin{Theorem} \label{thmCapacityMMNwithAWGN}
 Let $(\sigma_{(i,j)}^2:(i,j)\in \mathcal{E})$ be a MMN with independent AWGNs and zero-delay edges under multicast demand $(\mathcal{V}, \mathcal{D})$, and let
 \begin{equation}
 \mathcal{S}(P_\mathcal{E}, P_\mathcal{I}, P)\triangleq \left\{(S_{(1,1)}, S_{(1, 2)}, \ldots, S_{(N,N)})\in \mathbb{R}_+^{N^2} \left|\, \parbox[c]{2.1 in}{$ \sum_{i=1}^N\sum_{j=1}^N S_{(i,j)}\le P,\\
 \sum_{j=1}^N S_{(i,j)} \le P_i \text{ for all $i\in \mathcal{I}$,}
  \\
  S_{(i,j)} \le P_{(i,j)}  \text{ for all $(i,j)\in \mathcal{E}$}
 $}\right.\right\}\label{defSetS}
 \end{equation}
 be a set of $N^2$-dimensional tuples which specify the power allocation for the edges in the network. Let $S_{\mathcal{E}}$ denote $(S_{(1,1)}, S_{(1, 2)}, \ldots, S_{(N,N)})$ and define
 \begin{align}
\mathcal{R}_{\text{cut-set}} \triangleq \!\!\!\!\! \bigcup_{S_{\mathcal{E}}\in \mathcal{S}(P_\mathcal{E}, P_\mathcal{I}, P)} \bigcap_{T\subseteq \mathcal{I}: T^c \cap \mathcal{D} \ne \emptyset } \left\{ R_\mathcal{I}\in \mathbb{R}_+^N\left| \,  \parbox[c]{2.3 in}{$\sum\limits_{i\in T}R_i  \le \sum\limits_{(i,j)\in T\times T^c} \frac{1}{2}\log\Big(1+\frac{S_{(i,j)}}{\sigma_{(i,j)}^2}\Big)$} \right. \right\}\label{RoutAWGN}
\end{align}
to be the classical cut-set bound. Then, we have
 \[
\mathcal{C}=\mathcal{C}^{\mathbf{1}}=\mathcal{R}_{\text{cut-set}},
\]
where $\mathcal{C}$ and $\mathcal{C}^{\mathbf{1}}$ denote the capacity region and the~$\mathbf{1}$-capacity region respectively.
\end{Theorem}
\begin{Remark}
Theorem~\ref{thmCapacityMMNwithAWGN} generalizes the capacity result in Theorem~\ref{thmMMNwithIndDMCs} to the MMN consisting of independent AWGNs with zero-delay edges. Although the extra power constraints in \eqref{ineqnPij}, \eqref{ineqnPi} and \eqref{ineqnP} are introduced in the AWGN channels setting compared with the DMCs setting, we can still show in Theorem~\ref{thmCapacityMMNwithAWGN} that the set of achievable rate tuples for the MMN with independent AWGNs does not depend on the amount of delay incurred by each edge on each other edge even when zero-delay edges are present.
\end{Remark}

\subsection{Achievability}
Our goal is to prove
\begin{equation}
\mathcal{C}\supseteq \mathcal{C}^{\mathbf{1}} \supseteq \mathcal{R}_{\text{cut-set}}. \label{finalGoalInAchAWGN}
\end{equation}
Since $\mathcal{C}\supseteq \mathcal{C}^{\mathbf{1}}$ by \eqref{defCapacityRegionAWGN}, it suffices to show $\mathcal{C}^{\mathbf{1}} \supseteq \mathcal{R}_{\text{cut-set}}$, which is equivalent to the following statement by \eqref{RoutAWGN}:
\begin{equation}
\mathcal{C}^{\mathbf{1}}\supseteq \bigcap_{T\subseteq \mathcal{I}: T^c \cap \mathcal{D} \ne \emptyset } \left\{ R_\mathcal{I}\in \mathbb{R}_+^N\left| \,  \parbox[c]{2.6 in}{$\sum\limits_{i\in T}R_i  \le \sum\limits_{(i,j)\in T\times T^c} \frac{1}{2}\log\Big(1+\frac{(S_{(i,j)}^*-\delta)_+}{\sigma_{(i,j)}^2}\Big)$} \right. \right\}  \label{goalInAchAWGN}
\end{equation}
holds for all $S_\mathcal{E}^*\in  \mathcal{S}(P_\mathcal{E}, P_\mathcal{I}, P)$ and all $\delta>0$. In order to show \eqref{goalInAchAWGN}, we fix an $S_\mathcal{E}^*\in  \mathcal{S}(P_\mathcal{E}, P_\mathcal{I}, P)$ and a $\delta>0$. Suppose we use a random Gaussian codebook with power $S_{(i,j)}^*-\delta$ for each edge~$(i,j)\in \mathcal{E}$ so that the rate
 \begin{equation}
\mathrm{C}(S_{(i,j)}^*-\delta)\triangleq \frac{1}{2}\log\left(1+\frac{(S_{(i,j)}^*-\delta)_+}{\sigma_{(i,j)}^2}\right) \label{eqnAWGNCapacity}
\end{equation}
can be achieved for the AWGN $q_{Y_{(i,j)}|X_{(i,j)}}$ (defined in \eqref{defChannelMMNindAWGNs}) as~$n$ tends to infinity. In addition,
the power constraints \eqref{ineqnPij}, \eqref{ineqnPi} and \eqref{ineqnP} hold with probability approaching~$1$ by the weak law of large numbers due to the random Gaussian codebooks and the fact that $S_\mathcal{E}^*\in\mathcal{S}(P_\mathcal{E}, P_\mathcal{I}, P)$ (cf.\ \eqref{defSetS}). In the rest of the proof, we follow the network equivalence and NNC arguments in Section~\ref{subsectionAchIndDMCs} which have been used for the achievability proof of Theorem~\ref{thmMMNwithIndDMCs}.

Consider a deterministic counterpart of the MMN consisting of independent AWGNs by replacing every AWGN $q_{Y_{(i,j)}|X_{(i,j)}}$ with a noiseless bit pipe whose capacity is equal to $\mathrm{C}(S_{(i,j)}^*-\delta)$ (cf.\ \eqref{eqnAWGNCapacity}), and let $\mathcal{C}_{\text{det}}^{\mathbf{1}}$ be the capacity region of the deterministic counterpart network \textit{with unit-delay edges} where $\mathbf{1}$ denotes the all-one delay profile. Letting $\mathcal{C}^{\mathbf{1}}$ denote the $\mathbf{1}$-capacity region of the MMN consisting of independent AWGNs, we conclude by using the network equivalence theory \cite{networkEquivalencePartI} that
 \begin{equation}
 \mathcal{C}^{\mathbf{1}}\stackrel{\text{(a)}}{\supseteq}\mathcal{C}_{\text{det}}^{\mathbf{1}}, \label{CsupersetCdetAWGN}
 \end{equation}
 where (a) is not an equality because the capacity of every AWGN $q_{Y_{(i,j)}|X_{(i,j)}}$ is $\mathrm{C}(S_{(i,j)}^*)$ rather than $\mathrm{C}(S_{(i,j)}^*-\delta)$.
  Define
 \begin{equation}
 \mathcal{R}^{\mathbf{1}}\triangleq
\bigcap_{T\subseteq \mathcal{I}: T^c \cap \mathcal{D} \ne \emptyset } \left\{ R_\mathcal{I}\in \mathbb{R}_+^N\left| \,  \parbox[c]{2.1 in}{$\sum\limits_{i\in T}R_i  \le \sum\limits_{(i,j)\in T\times T^c} \mathrm{C}(S_{(i,j)}^*-\delta)$} \right. \right\} \label{defR1achAWGN}
  \end{equation}
  to be the cut-set outer bound for the deterministic counterpart network. It follows from the NNC inner bound in \cite[Sec.~II-A]{Lim11} that $\mathcal{C}_{\text{det}}^{\mathbf{1}}\supseteq\mathcal{R}^{\mathbf{1}}$, which implies from \eqref{CsupersetCdetAWGN} that
 $
 \mathcal{C}^{\mathbf{1}}\supseteq \mathcal{R}^{\mathbf{1}}$ holds, 
 which then implies from~\eqref{defR1achAWGN} and~\eqref{eqnAWGNCapacity} that~\eqref{goalInAchAWGN} holds. Since \eqref{goalInAchAWGN} holds for all $S_\mathcal{E}^*\in  \mathcal{S}(P_\mathcal{E}, P_\mathcal{I}, P)$ and all $\delta>0$, \eqref{finalGoalInAchAWGN} also holds.
\subsection{Converse}
Our goal is to prove
\begin{equation}
\mathcal{C}\subseteq\mathcal{R}_{\text{cut-set}}.  \label{finalGoalInConverseAWGN}
\end{equation}
 It suffices to show that
\begin{equation}
\mathcal{C}^{\boldsymbol{\pi}}\subseteq\mathcal{R}_{\text{cut-set}}  \label{goalInConverseAWGN}
\end{equation}
holds for all channel operation sequence $\boldsymbol{\pi}$,
which will then imply from \eqref{defCapacityRegionAWGN} that
\eqref{finalGoalInConverseAWGN} holds. Fix a channel operation sequence $\boldsymbol{\pi}$. In order to show \eqref{goalInConverseAWGN}, we fix an arbitrary delay profile~$B$ that is feasible with respect to~$\boldsymbol{\pi}$ and let $R_\mathcal{I}$ be a $(\boldsymbol{\pi},B)$-achievable rate tuple. By definition, there exists a sequence of $(\boldsymbol{\pi},B, n, M_{\mathcal{I}}, P_\mathcal{E}, P_\mathcal{I}, P)$-codes such that
\begin{equation}
 \liminf_{n\rightarrow \infty} \frac{\log M_{i}}{n} \ge R_{i} \label{eqnRateLimitAWGN}
\end{equation}
for each $i\in\mathcal{I}$ and
\begin{equation}
 \lim_{n\rightarrow \infty} P_{\text{err}}^{n} = 0.  \label{eqnProbLimitAWGN}
\end{equation}
where $P_{\text{err}}^{n}$ is as defined in~\eqref{defErrorProbabilitySt}.
Fix~$n$ and the corresponding $(\boldsymbol{\pi}, B, n, M_{\mathcal{I}}, P_\mathcal{E}, P_\mathcal{I}, P)$-code, and let $p_{W_\mathcal{I}, X_\mathcal{E}^n, Y_\mathcal{E}^n, \hat W_{\mathcal{I}\times \mathcal{I}}}$ be the probability distribution induced by the code. Fix any $T\subseteq \mathcal{I}$ such that $T^c\cap \mathcal{D}\ne \emptyset$. Following similar procedures for deriving \eqref{cutsetStatement5*} in Section~\ref{sectionInnerOuterBound}, we obtain from Fano's inequality, Proposition~\ref{propositionXFunctionOfY} and \eqref{memorylessStatementAWGN} that
\begin{align}
  & \left(1-P_{\text{err}}^{n}\right)\sum_{i\in T}\frac{1}{n}\log M_i
 \notag\\
 &\quad \le \frac{1}{n}\sum_{k=1}^n \sum_{h:\Omega_{h}\in \mathcal{I}\times T^c}  \big(h_{p_{X_\mathcal{E}^n, Y_\mathcal{E}^n}}(Y_{(\mathcal{I}\times T^c)\cap \Omega_{h} ,k}| X_{(T^c \times \mathcal{I})\cap \Omega_{h},k}) - h_{p_{X_\mathcal{E}^n, Y_\mathcal{E}^n}}(Y_{(\mathcal{I}\times T^c)\cap \Omega_{h} ,k}| X_{\Omega_{h},k})\big). \label{cutsetStatement4AWGN}
\end{align}
Following \eqref{cutsetStatement4AWGN}, we consider the following chain of equalities for each $k\in\{1, 2, \ldots, n\}$:
\begin{align}
&  \sum_{h:\Omega_{h}\in \mathcal{I}\times T^c} \big(h_{p_{X_\mathcal{E}^n, Y_\mathcal{E}^n}}(Y_{(\mathcal{I}\times T^c)\cap \Omega_{h} ,k}| X_{(T^c \times \mathcal{I})\cap \Omega_{h},k}) - h_{p_{X_\mathcal{E}^n, Y_\mathcal{E}^n}}(Y_{(\mathcal{I}\times T^c)\cap \Omega_{h} ,k}| X_{\Omega_{h},k})\big)\notag\\
 &\quad \stackrel{\text{(a)}}{=} \sum_{(i,j)\in \mathcal{I}\times T^c} \big(h_{p_{X_\mathcal{E}^n, Y_\mathcal{E}^n}}(Y_{(\mathcal{I}\times T^c)\cap \{(i,j)\} ,k}| X_{(T^c \times \mathcal{I})\cap \{(i,j)\},k}) - h_{p_{X_\mathcal{E}^n, Y_\mathcal{E}^n}}(Y_{(\mathcal{I}\times T^c)\cap \{(i,j)\} ,k}| X_{(i,j),k})\big)\notag\\
 &\quad  \stackrel{\text{(b)}}{=} \sum_{(i,j)\in T\times T^c} \big(h_{p_{X_\mathcal{E}^n, Y_\mathcal{E}^n}}(Y_{(\mathcal{I}\times T^c)\cap \{(i,j)\} ,k}) - h_{p_{X_\mathcal{E}^n, Y_\mathcal{E}^n}}(Y_{(\mathcal{I}\times T^c)\cap \{(i,j)\} ,k}| X_{(i,j),k})\big)\notag\\
 &\quad = \sum_{(i,j)\in T\times T^c} \big(h_{p_{X_\mathcal{E}^n, Y_\mathcal{E}^n}}(Y_{(i,j) ,k}) - h_{p_{X_\mathcal{E}^n, Y_\mathcal{E}^n}}(Y_{(i,j) ,k}| X_{(i,j),k})\big), \label{cutsetStatement5AWGN}
\end{align}
where
\begin{enumerate}
\item[(a)] follows from the facts that $(\Omega_{\pi(1)}, \Omega_{\pi(2)}, \ldots, \Omega_{\pi(N^2)})$ is an~$N^2$ partition of $\mathcal{E}$ and each $\Omega_{h}$ contains exactly one edge.
    \item[(b)] follows from the fact that for each $(i,j) \in T^c\times T^c$,
    \[
    h_{p_{X_\mathcal{E}^n, Y_\mathcal{E}^n}}(Y_{(\mathcal{I}\times T^c)\cap \{(i,j)\} ,k}| X_{(T^c \times \mathcal{I})\cap \{(i,j)\},k}) =  h_{p_{X_\mathcal{E}^n, Y_\mathcal{E}^n}}(Y_{(\mathcal{I}\times T^c)\cap \{(i,j)\} ,k}| X_{(i,j),k}).
    \]
\end{enumerate}
Let $p_{Z_{(i,j),k}}$ be the distribution of the zero-mean Gaussian random variable $Z_{(i,j),k}$ whose variance is $\sigma_{(i,j)}^2$ for each $(i,j)\in \mathcal{E}$ and each $k\in\{1,2, \ldots, n\}$. It then follows from \eqref{memorylessStatementAWGN*} and \eqref{defChannelMMNindAWGNs} that $Y_{(i,j),k}$ has the same distribution as $X_{(i,j),k}+Z_{(i,j),k}$ when $Y_{(i,j),k}$ and $(X_{(i,j),k},Z_{(i,j),k})$ are distributed according to $p_{X_\mathcal{E}^n, Y_\mathcal{E}^n}$ and $p_{X_{(i,j),k}}p_{Z_{(i,j),k}}$ respectively for each $(i,j)\in \mathcal{E}$ and each $k\in\{1,2, \ldots, n\}$, which implies that
 \begin{align}
&\frac{1}{n}\sum_{k=1}^n \sum_{(i,j)\in T\times T^c} \big(h_{p_{X_\mathcal{E}^n, Y_\mathcal{E}^n}}(Y_{(i,j) ,k}) - h_{p_{X_\mathcal{E}^n, Y_\mathcal{E}^n}}(Y_{(i,j) ,k}| X_{(i,j),k})\big)\notag\\
& = \frac{1}{n}\sum_{k=1}^n \sum_{(i,j)\in T\times T^c} \big(h_{p_{X_{(i,j),k}}p_{Z_{(i,j),k}}}(X_{(i,j) ,k}+Z_{(i,j) ,k}) - h_{p_{X_{(i,j),k}}p_{Z_{(i,j),k}}}(X_{(i,j) ,k}+Z_{(i,j) ,k}| X_{(i,j),k})\big). \label{cutsetStatement6AWGN}
 \end{align}
Following \eqref{cutsetStatement6AWGN}, we consider
 \begin{align}
 h_{p_{X_{(i,j),k}}p_{Z_{(i,j),k}}}(X_{(i,j) ,k}+Z_{(i,j) ,k})&  \stackrel{\text{(a)}}{\le} \log \sqrt{2\pi e \left(\Var_{p_{X_{(i,j),k}}}\left[X_{(i,j),k}\right]+\Var_{p_{Z_{(i,j),k}}}\left[Z_{(i,j),k}\right]\right)} \notag\\
 & \le \log \sqrt{2\pi e \left(\E_{p_{X_{(i,j),k}}}\left[X_{(i,j),k}^2\right]+\Var_{p_{Z_{(i,j),k}}}\left[Z_{(i,j),k}\right]\right)} \notag\\
 & =  \log \sqrt{2\pi e \left(\E_{p_{X_{(i,j),k}}}\left[X_{(i,j),k}^2\right]+\sigma_{(i,j)}^2\right)} \label{cutsetStatement7AWGN}
 \end{align}
 and
  \begin{align}
h_{p_{X_{(i,j),k}}p_{Z_{(i,j),k}}}(X_{(i,j) ,k}+Z_{(i,j) ,k}| X_{(i,j),k})&  = h_{p_{Z_{(i,j),k}}}(Z_{(i,j) ,k}) \notag\\
 & = \log \sqrt{2\pi e \Var_{p_{Z_{(i,j),k}}}\left[Z_{(i,j),k}\right]}\notag\\
 & = \log \sqrt{2\pi e \sigma_{(i,j)}^2} \label{cutsetStatement8AWGN}
 \end{align}
for each $(i,j)\in \mathcal{E}$ and $k\in\{1, 2, \ldots, n\}$,
where (a) follows from the fact that the differential entropy of a random variable~$X$ is maximized by that of the zero-mean Gaussian random variable whose variance is $\Var[X]$. Combining \eqref{cutsetStatement4AWGN}, \eqref{cutsetStatement5AWGN}, \eqref{cutsetStatement6AWGN}, \eqref{cutsetStatement7AWGN} and \eqref{cutsetStatement8AWGN}, we have
\begin{align}
 \left(1-P_{\text{err}}^{n}\right)\sum_{i\in T}\frac{1}{n}\log M_i
 & \le \frac{1}{n}\sum_{k=1}^n \sum_{(i,j)\in T\times T^c} \frac{1}{2}\log\left(1+\frac{\E_{p_{X_{(i,j),k}}}\left[X_{(i,j),k}^2\right]}{\sigma_{(i,j)}^2}\right)
\notag\\
&  \stackrel{\text{(a)}}{\le}  \sum_{(i,j)\in T\times T^c} \frac{1}{2}\log\left(1+\frac{\E_{p_{X_{(i,j)}^n}}\left[\frac{1}{n}\sum_{k=1}^n X_{(i,j),k}^2\right]}{\sigma_{(i,j)}^2}\right) \label{cutsetStatement9AWGN}
\end{align}
where (a) follows from Jensen's inequality. Choose $\{n_\ell\}_{\ell=1}^{\infty}$ to be a subsequence of $\{n\}_{n=1}^\infty$ such that for all $(i,j)\in \mathcal{E}$, $\lim\limits_{\ell\rightarrow \infty} \E_{p_{X_{(i,j)}^{n_\ell}}}\left[ \frac{1}{n_\ell}\sum_{k=1}^{n_\ell} X_{(i,j),k}^2\right]$ converges to some $S_{(i,j)}^*\in \mathbb{R}$, i.e.,
\begin{align}
S_{(i,j)}^*=\lim\limits_{\ell\rightarrow \infty} \E_{p_{X_{(i,j)}^{n_\ell}}}\left[ \frac{1}{n_\ell}\sum_{k=1}^{n_\ell} X_{(i,j),k}^2\right] . \label{cutsetStatement10AWGN}
\end{align}
Combining \eqref{eqnRateLimitAWGN}, \eqref{eqnProbLimitAWGN}, \eqref{cutsetStatement9AWGN} and \eqref{cutsetStatement10AWGN}, we obtain
\begin{equation}
\sum_{i\in T}R_i \le \sum_{(i,j)\in T\times T^c} \frac{1}{2}\log\left(1+\frac{S_{(i,j)}^*}{\sigma_{(i,j)}^2}\right). \label{cutsetStatement11AWGN}
\end{equation}
Define $S_\mathcal{E}^*\triangleq (S_{(i,j)}^*: (i,j)\in \mathcal{E})$ to be an~$N^2$ dimensional-tuple. Since $S_\mathcal{E}^*$ does not depend on~$T$ by \eqref{cutsetStatement10AWGN}, it follows from \eqref{cutsetStatement11AWGN} that
 \begin{align}
R_\mathcal{I} \in \bigcap_{T\subseteq \mathcal{I}: T^c \cap \mathcal{D} \ne \emptyset } \left\{ R_\mathcal{I}\in \mathbb{R}_+^N\left| \,  \parbox[c]{2.3 in}{$\sum\limits_{i\in T}R_i  \le \sum\limits_{(i,j)\in T\times T^c} \frac{1}{2}\log\Big(1+\frac{S_{(i,j)}^*}{\sigma_{(i,j)}^2}\Big)$} \right. \right\}.\label{cutsetStatement12AWGN}
 \end{align}
On the other hand, we conclude from \eqref{cutsetStatement10AWGN}, \eqref{ineqnPij}, \eqref{ineqnPi} and \eqref{ineqnP} that
\begin{equation}
S_{(i,j)}^* \le P_{(i,j)} \label{ineqnSij}
\end{equation}
for each $(i,j)\in \mathcal{E}$,
\begin{equation}
\sum_{j=1}^N S_{(i,j)}^* \le P_i \label{ineqnSi}
\end{equation}
for each $i\in \mathcal{I}$, and
\begin{equation}
\sum_{i=1}^N\sum_{j=1}^N S_{(i,j)}^* \le P. \label{ineqnS}
\end{equation}
Combining \eqref{defSetS}, \eqref{ineqnSij}, \eqref{ineqnSi} and \eqref{ineqnS}, we have
\begin{equation}
S_\mathcal{E}^* \in \mathcal{S}(P_\mathcal{E}, P_\mathcal{I}, P). \label{eqnSE*inS}
\end{equation}
Consequently, \eqref{goalInConverseAWGN} follows from \eqref{cutsetStatement12AWGN} and~\eqref{eqnSE*inS}. Since \eqref{goalInConverseAWGN} holds for all channel operation sequence $\boldsymbol{\pi}$, \eqref{finalGoalInConverseAWGN} also holds.

\section{Concluding Remarks} \label{sectionConclusion}
We investigate the MMN consisting of independent channels and propose an edge-delay model which allows the presence of zero-delay edges in the network. Under our model, the MMN with independent channels is characterized by multiple channels, which are operated in different orders so that an edge may incur zero delay on some other edges. Our model is a generalization of the classical model, under which the MMN with independent channels is characterized by a single channel and every edge incurs a unit delay on every adjacent edge. A well-known outer bound on the capacity region under the classical unit-delay assumption is the cut-set outer bound. In this paper, we prove that the MMN with independent channels and zero-delay edges lies within the classical cut-set bound despite a violation of the classical unit-delay assumption.

Next, we use our outer bound to prove the capacity region of the MMN consisting of independent DMCs with zero-delay edges. More specifically, we show that the capacity region is the same as the set of achievable rate tuples under the classical unit-delay assumption. This capacity region result is then generalized to the MMN consisting of independent AWGN channels with zero-delay edges. Consequently, the capacity regions of the two aforementioned MMNs are not affected by the handling of delays among edges even when zero-delay edges are present.

It has been shown in Section~\ref{sectionMotivating} that the capacity region of some two-node MMN with \textit{dependent} channels and zero-delay edges is strictly larger than the set of achievable rate tuples under the classical unit-delay assumption --- every edge should incur a unit delay on all the edges. Future research may investigate the capacity regions for general MMNs consisting of \textit{dependent} channels with zero-delay edges. Another interesting direction for future research is extending the network equivalence theory~\cite{networkEquivalencePartI} (which asserts the equivalence between the capacity region of any network with independent channels and the capacity region of the deterministic counterpart of the network) for networks with unit-delay edges to networks with zero-delay edges.

\appendices



%
%



\end{document}